
\def\singlespace{\normalbaselines}
\def\oneandahalfspace{\baselineskip=16pt plus 1pt
\lineskip=2pt\lineskiplimit=1pt}

\def\np{\vfill\eject}
\def\nl{\hfil\break}

\def\nofirstpagenoten{\nopagenumbers\footline={\ifnum\pageno>1\tenrm
\hss\folio\hss\fi}}
\def\nofirstpagenotwelve{\nopagenumbers\footline={\ifnum\pageno>1\twelverm
\hss\folio\hss\fi}}
\def\leaderfill{\leaders\hbox to 1em{\hss.\hss}\hfill}


\parindent=20pt
\def\narrow{\advance\leftskip by 40pt \advance\rightskip by 40pt}

\def\AB{\bigskip
        \centerline{\bf ABSTRACT}\medskip\narrow}
\def\nonarrower{\advance\leftskip by -40pt\advance\rightskip by -40pt}
\def\AE{\bigskip\nonarrower}

\def\boxit#1{\vbox{\hrule\hbox{\vrule\kern3pt
        \vbox{\kern3pt#1\kern3pt}\kern3pt\vrule}\hrule}}

\def\gtorder{\mathrel{\raise.3ex\hbox{$>$}\mkern-14mu
             \lower0.6ex\hbox{$\sim$}}}
\def\ltorder{\mathrel{\raise.3ex\hbox{$<$}|mkern-14mu
             \lower0.6ex\hbox{\sim$}}}
\def\dalemb#1#2{{\vbox{\hrule height .#2pt
        \hbox{\vrule width.#2pt height#1pt \kern#1pt
                \vrule width.#2pt}
        \hrule height.#2pt}}}

\font\twelvett=cmtt12 \font\twelvebf=cmbx12
\font\twelverm=cmr12 \font\twelvei=cmmi12 \font\twelvess=cmss12
\font\twelvesy=cmsy10 scaled \magstep1 \font\twelvesl=cmsl12
\font\twelveex=cmex10 scaled \magstep1 \font\twelveit=cmti12
\font\tenss=cmss10
 
 \font\ninebf=cmbx9
\font\ninerm=cmr9 \font\ninei=cmmi9
\font\ninesy=cmsy9 
\font\eightrm=cmr8
\catcode`@=11
\newskip\ttglue
\newfam\ssfam

\def\twelvepoint{\def\rm{\fam0\twelverm}
\textfont0=\twelverm \scriptfont0=\ninerm \scriptscriptfont0=\sevenrm
\textfont1=\twelvei \scriptfont1=\ninei \scriptscriptfont1=\seveni
\textfont2=\twelvesy \scriptfont2=\ninesy \scriptscriptfont2=\sevensy
\textfont3=\twelveex \scriptfont3=\twelveex \scriptscriptfont3=\twelveex
\def\it{\fam\itfam\twelveit} \textfont\itfam=\twelveit
\def\sl{\fam\slfam\twelvesl} \textfont\slfam=\twelvesl
\def\bf{\fam\bffam\twelvebf} \textfont\bffam=\twelvebf
\scriptfont\bffam=\ninebf \scriptscriptfont\bffam=\sevenbf
\def\tt{\fam\ttfam\twelvett} \textfont\ttfam=\twelvett
\def\ss{\fam\ssfam\twelvess} \textfont\ssfam=\twelvess
\tt \ttglue=.5em plus .25em minus .15em
\normalbaselineskip=14pt
\abovedisplayskip=14pt plus 3pt minus 10pt
\belowdisplayskip=14pt plus 3pt minus 10pt
\abovedisplayshortskip=0pt plus 3pt
\belowdisplayshortskip=8pt plus 3pt minus 5pt
\parskip=3pt plus 1.5pt
\setbox\strutbox=\hbox{\vrule height10pt depth4pt width0pt}
\let\sc=\ninerm
\let\big=\twelvebig \normalbaselines\rm}
\def\twelvebig#1{{\hbox{$\left#1\vbox to10pt{}\right.\n@space$}}}

\def\tenpoint{\def\rm{\fam0\tenrm}
\textfont0=\tenrm \scriptfont0=\sevenrm \scriptscriptfont0=\fiverm
\textfont1=\teni \scriptfont1=\seveni \scriptscriptfont1=\fivei
\textfont2=\tensy \scriptfont2=\sevensy \scriptscriptfont2=\fivesy
\textfont3=\tenex \scriptfont3=\tenex \scriptscriptfont3=\tenex
\def\it{\fam\itfam\tenit} \textfont\itfam=\tenit
\def\sl{\fam\slfam\tensl} \textfont\slfam=\tensl
\def\bf{\fam\bffam\tenbf} \textfont\bffam=\tenbf
\scriptfont\bffam=\sevenbf \scriptscriptfont\bffam=\fivebf
\def\tt{\fam\ttfam\tentt} \textfont\ttfam=\tentt
\def\ss{\fam\ssfam\tenss} \textfont\ssfam=\tenss
\tt \ttglue=.5em plus .25em minus .15em
\normalbaselineskip=12pt
\abovedisplayskip=12pt plus 3pt minus 9pt
\belowdisplayskip=12pt plus 3pt minus 9pt
\abovedisplayshortskip=0pt plus 3pt
\belowdisplayshortskip=7pt plus 3pt minus 4pt
\parskip=0.0pt plus 1.0pt
\setbox\strutbox=\hbox{\vrule height8.5pt depth3.5pt width0pt}
\let\sc=\eightrm
\let\big=\tenbig \normalbaselines\rm}
\def\tenbig#1{{\hbox{$\left#1\vbox to8.5pt{}\right.\n@space$}}}
\let\rawfootnote=\footnote \def\footnote#1#2{{\rm\parskip=0pt\rawfootnote{#1}
{#2\hfill\vrule height 0pt depth 6pt width 0pt}}}
\def\tenfoot{\tenpoint\hskip-\parindent\hskip-.1cm}

\overfullrule=0pt
\twelvepoint
\def\sbullet{\raise.2em\hbox{$\scriptscriptstyle\bullet$}}
\nofirstpagenotwelve
\hsize=16.5 truecm
\baselineskip 15pt

\def\ft#1#2{{\textstyle{{#1}\over{#2}}}}

\def\a{\alpha}

\def\del{\partial}
\def\df{\del\varphi}

\def\w24{$W_{2,4}$}

\oneandahalfspace
\rightline{CTP TAMU--70/93}
\rightline{LZU-TH-93/09}
\rightline{hep-th/9311084}
\rightline{November 1993}

\vskip 2truecm
\centerline{\bf Critical and Non-critical $W_{2,4}$ Strings}
\vskip 1.5truecm
\centerline{H. Lu, C.N. Pope,\footnote{$^*$}{\tenfoot \sl  Supported in
part by the U.S. Department of Energy, under
grant DE-FG05-91ER40633.}
X.J.~Wang and
S.C Zhao\footnote{$^\dagger$}{\tenfoot \sl
On leave of absence from Lanzhou University, P.R.~China}}
\vskip 1.5truecm
\centerline{\it Center for Theoretical Physics,
Texas A\&M University,}
\centerline{\it College Station, TX 77843--4242, USA.}

\vskip 1.5truecm
\AB\singlespace
 Nilpotent BRST operators for higher-spin $W_{2,s}$ strings, with currents of
spins 2 and $s$, have recently been constructed for $s=4$, 5 and 6.  In the
case
of $W_{2,4}$, this operator can be understood as being the BRST operator for
the critical $W\!B_2$ string.  In this paper, we construct a generalised BRST
operator that can be associated with a non-critical $W_{2,4}$ string, in which
$W\!B_2$ matter is coupled to the $W\!B_2$ gravity of the critical case.  We
also obtain the complete cohomology of the critical $W_{2,4}$ BRST operator,
and
investigate the physical spectra of the $s=5$ and $s=6$ string theories.
\AE
\oneandahalfspace

\np
\noindent
{\bf 1. Introduction}
\bigskip

     There has been much progress in understanding $W$-string theories
recently.
$W$ strings are two-dimensional quantum field theories with non-linear local
symmetry algebras, namely $W$ algebras, which are the higher-spin
generalisations of the Virasoro algebra.   There exist many higher-spin
extensions of the Virasoro algebra, and the corresponding string theories are
interesting subjects which may help us better to understand the fundamental
properties of two-dimensional quantum field theories. There are various
approaches to the quantisation of the usual bosonic string since the symmetry
algebra, namely the Virasoro algebra, is linear.  However, since $W$ algebras
are generically non-linear and a physical gauge seems not to exist, BRST
methods appear to provide the only viable approach to the quantisation of
$W$-string theories.  The characteristics of the $W$ string are therefore
governed by the structure of the corresponding nilpotent BRST operator.   Since
$W$ algebras are non-linear, BRST operators can be complicated to construct,
and
in fact there may exist intrinsically different nilpotent BRST operators for a
given $W$ algebra, which give rise to different $W$-string theories.

     The simplest example of a $W$ algebra is the $W_3$ algebra of
Zamolodchikov [1].   The algebra contains a spin-$3$ primary current $W$ in
addition to the spin-$2$ energy-momentum tensor $T$.  The first requirement of
building a $W_3$ string is an anomaly-free theory of $W_3$ gravity.   In [2],
it
was shown that such a theory could be obtained by standard BRST techniques. The
result then leads on to the construction of a $W_3$ string [3,4,5].    A BRST
operator for $W_3$ was first constructed in [6]; its nilpotency  demands that
the matter currents $T$ and $W$ generate the $W_3$ algebra with central charge
$c=100$ [6,7].  The detailed study of the $W_3$ string based on this BRST
operator can be found in [5,8,9,10,11].   This $W_3$ string is sometimes called
the ``critical'' $W_3$ string, for reasons that will be clarified presently.

      In [12,13], a BRST operator for the so-called non-critical $W_3$ string
was found. The non-critical $W_3$ string is a theory of $W_3$ matter coupled to
$W_3$ gravity; it is a generalisation of two-dimensional matter coupled to
gravity. The BRST operator of the non-critical $W_3$ string contains two
copies of the $W_3$ algebra. One is the $W$-gravity, or Toda, contribution,
which generalises the Liouville gravity sector of the non-critical bosonic, and
the other is the matter contribution. That such a tensoring of two copies of
the
$W_3$ algebra can give rise to a sensible theory is quite non-trivial since the
$W_3$ algebra is non-linear. The nilpotency of the BRST operator requires the
total central charge of the Toda and matter sectors together to be $c=100$.

     Although we are following the conventional terminology of [12], and using
the term ``critical'' to describe the original BRST operator of [6] involving
only one copy of the $W_3$ algebra, whilst using the term ``non-critical'' to
describe the BRST operator of [12,13] that involves two commuting copies of the
$W_3$ algebra, it should be emphasised that both BRST operators are in
fact critical in the sense that they are nilpotent operators
built from a set of fields satisfying free-field OPEs.  In the case of the
ordinary bosonic string the term ``non-critical'' tends to be reserved for the
situation where one of the fields (the Liouville field) arises dynamically,
with an exponential potential term, as a consequence of the Weyl anomaly
associated with the non-nilpotence of the BRST operator built just from the
ghosts and the matter fields.  Although the ``non-critical'' $W_3$ BRST
operator
of [12,13] might provide the appropriate arena for studying the analogous
emergence of dynamical Toda fields in $W_3$ gravity coupled to $W_3$ matter, in
all of the discussions to date it has been used only as a description of a
critical $W_3$ string theory with free-field quantisation rules.  Further
discussion of this point may be found in [14].  Even in the absence of the
exponential potential terms, the ``non-critical'' $W_3$ BRST operator can be,
and indeed is, different from the ``critical'' one.  This can happen because,
unlike the usual Virasoro case, the algebra is non-linear and so different
kinds of BRST operator can occur.

      A natural generalisation of the $W_3$-string is a higher-spin string with
local spin-2 and spin-$s$ symmetries on the world-sheet, instead of spin-2 and
spin-3. Critical BRST operators for such theories have been constructed for
$s=4,5,6$ [15].  Presumably they exist for all $s$, although the complexity
rises  rapidly with increasing $s$.  In fact, it was found in  [15] that there
are two different BRST operators when $s=4$, one operator  when $s=5$, and four
different ones when $s=6$.  For each value of $s$, it  appears that one of the
BRST operators is associated with a unitary  multi-scalar string theory [15].
It is these BRST operators with which we  shall be concerned in this paper.  We
shall refer to the associated gauge theories as $W_{2,s}$ strings.  The
physical
spectrum of the multi-scalar critical $W_{2,s}$ string is closely related to
the
lowest unitary $W_{(s-1)}$ minimal model [15,16].

     The $W_{2,4}$ BRST operator is in fact a BRST operator for the $W\!B_2$
algebra.  However, it should be remarked that for the higher $W_{2,s}$ BRST
operators, there does not necessarily exist a corresponding closed $W_{2,s}$
algebra at the quantum level.  For example, a closed algebra of spin-2 and
spin-5 currents exists only at certain discrete values of central charge
[17,18], which do not include the value that would be needed for
criticality.   Although a closed algebra of spin-2 and spin-6 currents,
namely $W\!G_2$,  exists for all values of central charge [19], the
$W_{2,6}$ string that we are considering is not related to this algebra. (In
fact two of the other three BRST operators with $s=6$ constructed in [15] are
related to the $W\!G_2$ algebra.)  For all $s>6$, it is known that closed
algebras  of spin-2 and spin-$s$ currents could exist for at most a finite
number of central-charge values, which would presumably not include the
critical
values.

     In this paper, we begin in section 2 by reviewing the critical \w24 string
theory that was obtained in [15].  Then, in section 3, we construct a
non-critical BRST  operator for the \w24 string, generalising the $W_3$ results
of [12,13,20].  We use  this in section 4 to obtain some of the physical
states,
including some of the  ghost-number zero ground-ring generators.  It would be
interesting to  determine the complete cohomologies of the non-critical and
critical \w24 strings along the lines of the results obtained in [21,11] for
the
$W_3$ string. For the non-critical case, this is a
difficult problem, and in the present paper, in section 5, we only
consider  the simpler critical case. Here, we may use the methods of [11] to
obtain the cohomology by acting with  powers of certain invertible physical
operators on a basic set of physical  states. A partial analysis for
higher-spin
$W_{2,s}$ strings in section 6 reveals interesting new features.

\bigskip\bigskip
\noindent
{\bf 2.  Review of the Critical $W_{2,s}$ String}
\bigskip

    It was shown in [10] that the BRST operator for the usual critical $W_3$
string could be brought into a simpler form by performing a non-linear
transformation under which the ghosts $(b,c)$ and $(\beta,\gamma)$ for the
spin-2 and spin-3 currents, and one of the matter fields, are mixed.  In terms
of the redefined fields, the BRST operator has a double grading, with respect
to
the ghost numbers for $(b,c)$ and $(\beta,\gamma)$ respectively.  In [15],  it
was shown that this graded BRST operator can be generalised to one where the
matter currents have spins 2 and $s$ rather than 2 and 3.  Note
that $\beta$ therefore has spin $s$, and $\gamma$ has spin $(1-s)$.  The BRST
operator for the spin-2 plus spin-$s$ string then takes the form [15]:
$$
\eqalignno{ Q_B&= Q_0 + Q_1,&(2.1)\cr
Q_0&=\oint dz\, c \Big(T_{\varphi_1}+T_{\varphi_2} + T_{\gamma,\beta} + \ft12\,
T_{c,b} \Big)\ , &(2.2)\cr
Q_1&=\oint dz\, \gamma \, F(\varphi_1,\beta,\gamma)\ ,&(2.3)\cr}
$$
where the energy-momentum tensors are given by
$$
\eqalignno{
T_{\varphi_1}&\equiv -\ft12 (\del\varphi_1)^2 -\alpha\, \del^2\varphi_1,
   &(2.4)\cr
T_{\varphi_2} &\equiv -\ft12 (\del \varphi_2)^2 - a\,\del^2 \varphi_2 \ ,
&(2.5)\cr
T_{\gamma,\beta}&\equiv -s\, \beta\,\del\gamma -(s-1)\, \del\beta\,
\gamma\ , &(2.6)\cr
T_{c,b}&\equiv -2\, b\, \del c - \del b\, c\ , &(2.7)\cr
}
$$
The operator $F(\varphi_1,\beta,\gamma)$ has spin $s$ and ghost number zero.
The
BRST operator is graded, as discussed above, with $Q_0^2=Q_1^2=\{Q_0,Q_1\}=0$.
The first of these conditions is satisfied provided that the total central
charge vanishes, {\it i.e.}\
$$
0=-26 -2(6s^2-6s+1) + 2+12\alpha^2 + 12a^2.\eqno(2.8)
$$
The remaining two nilpotency conditions determine the precise form of the
operator $F(\varphi_1,\beta,\gamma)$ appearing in (2.3).  Solutions for $s=4$,
5
and 6 were found in [15].  (In fact in [15] it was found that whilst there
is just one possible BRST operator for $s=5$, there are two different
BRST operators when $s=4$, and four different ones when $s=6$. However,
only one BRST operator for each $s$, for which the $(\varphi_1,\beta,\gamma)$
system has the central charge ${2(s-2)\over (s+1)}$ discussed above, seems to
be
associated with a unitary string theory.  It is this choice that we shall be
concentrating on in the present paper.)

     In this paper, we shall principally be interested in the case $s=4$.  The
BRST operator is then given by (2.1)--(2.7), with $\alpha^2=\ft{243}{20},\
a^2=\ft{121}{60}$, and [15]
$$
\eqalign{
F(\varphi_1,\beta,\gamma)&=(\del\varphi_1)^4 + 4\alpha\, \del^2\varphi_1\,
(\del\varphi_1)^2 + \ft{41}5 (\del^2\varphi_1)^2 + \ft{124}{15}
\del^3\varphi_1\, \del \varphi_1 +\ft{46}{135} \alpha\, \del^4\varphi_1\cr
& +8 (\del\varphi_1)^2\, \beta\, \del\gamma -\ft{16}9\alpha\, \del^2\varphi_1\,
\beta\, \del\gamma -\ft{32}9 \alpha\, \del\varphi_1\,
\beta\, \del^2 \gamma-\ft45 \beta\, \del^3\gamma + \ft{16}3 \del^2\beta\,
\del\gamma .\cr}\eqno(2.9)
$$
The results for $s=5$ and $s=6$ may be found in [15].

     Finally, we remark that the scalar field $\varphi_2$ appears in the BRST
operator only {\it via} its energy-momentum tensor (2.5), and thus
appears only in $Q_0$ but not $Q_1$.  Consequently, there exist more general
critical \w24 strings in which $T_{\varphi_2}$ is replaced by an arbitrary
energy-momentum tensor $T^{\rm eff}$ that has the same central charge as
$T_{\varphi_2}$, namely $c^{\rm eff}=\ft{126}5$, and that commutes with
$\varphi_1$.  In particular, one can take $T^{\rm eff}$ to be realised in terms
of a set of $d$ scalar fields  $X^\mu$, with a background charge, giving a
multi-scalar critical \w24 string [15].  This procedure can be applied to the
$W_{2,s}$ strings for higher values of $s$ also [15].

\bigskip\bigskip
\noindent
{\bf 3. The Non-critical $W_{2,4}$ String}
\bigskip

The non-critical \w24 string is a theory of \w24 gravity coupled to a
matter system on which the \w24 algebra is realised. The \w24 algebra
was constructed in [17,18] in terms of Laurent modes by imposing the Jacobi
identity. It is sometimes called $W\!B_2$, since it can be obtained by
Hamiltonian reduction from the (non-simply-laced) algebra $B_2$.
The algebra $W\!B_2$ can be re-expressed in terms of the spin-two and spin-four
currents $T$ and $W$.  Since $W$ is primary under $T$, the only OPE which we
need give is $W(z)W(w)$ [17,18]:
$$
\eqalign{
W(z)W(w)&\sim \Big\{ {2T\over (z-w)^6} + {\del T\over (z-w)^5} +
                 \ft{3}{10}{\del^2 T\over (z-w)^4}\cr
        &\phantom{00}+\ft{1}{15}{\del^3 T\over (z-w)^3} +
                 \ft{1}{84}{\del^4 T\over (z-w)^2} +
                 \ft{1}{560}{\del^5 T\over (z-w)} \Big \} \cr
        &+b_1\Big\{ {U\over (z-w)^4} + \ft12 {\del U\over (z-w)^3} +
                  \ft5{36}{\del^2 U\over (z-w)^2}+\ft1{36}{\del^3 U\over (z-w)}
            \Big \}\cr
        &+b_2\Big\{ {W\over (z-w)^4} + \ft12 {\del W\over (z-w)^3} +
                  \ft5{36}{\del^2 W\over (z-w)^2}+\ft1{36}{\del^3 W\over (z-w)}
            \Big\}\cr
        &+b_3\Big\{ {G\over (z-w)^2} + \ft12 {\del G\over (z-w)} \Big\} +
          b_4\Big\{ {A\over (z-w)^2} + \ft12 {\del A\over (z-w)} \Big\}\cr
        &+b_5\Big\{ {B\over (z-w)^2} + \ft12 {\del B\over (z-w)} \Big\}
         + {c/4\over (z-w)^8}\cr}
\eqno(3.1)
$$
where the (quasi-primary) composites $U$ (spin $4$), and $G, A, B$ (all spin
$6$), are defined by
$$
\eqalign{
U&\equiv (TT)-\ft3{10}\del^2 T\ ,\cr
G&\equiv (\del^2 T\, T)-\del (\del T\, T) + \ft{2}{9}\, \del^2(TT) -
\ft1{42}\, \del^4 T\ ,\cr
A&\equiv (TU)-\ft16\,\del^2 U\ ,\cr
B&\equiv (TW)-\ft16\,\del^2 W \ ,\cr}\eqno(3.2)
$$
with normal ordering of products of currents understood.   The coefficients
$b_1,b_2,b_3,b_4,b_5$ are given by
$$
\eqalign{
b_1&={42 \over 5c+22}\ ,\cr
b_2&={\sqrt { 54(c+24)(c^2-172c+196)\over (5c+22)(7c+68)(2c-1)}}\ ,\cr
b_3&={3 (19c-524)\over 10(7c+68)(2c-1)}\ ,\cr
b_4&={24(72c+13)\over (5c+22)(7c+68)(2c-1)}\ ,\cr
b_5&={28 \over 3(c+24)} b_2\ .\cr}
\eqno(3.3)
$$

     We are now in a position to present our results for the non-critical \w24
string.  The BRST operator takes the form
$$
\eqalignno{ Q_B&= Q_0 + Q_1,&(3.4)\cr
Q_0&=\oint dz\, c \Big(T_{\varphi_1}+ T_{\varphi_2}
+T_M+ T_{\gamma,\beta} + \ft12\,
T_{c,b} \Big), &(3.5)\cr
Q_1&=\oint dz\, \gamma \, F(\varphi_1,\beta,\gamma,T_M,W_M),&(3.6)\cr}
$$
where the matter currents $T_M$ and $W_M$ generate the $W\!B_2$ algebra, the
ghost energy-momentum tensors are given by (2.6) and (2.7), with $s=4$ in
(2.6),  $T_{\varphi_1}=-\ft12(\df_1)^2 -\alpha\, \del^2\varphi_1$, and
$T_{\varphi_2}=-\ft12(\df_2)^2 -a\, \del^2\varphi_2$. The
$\varphi_1$ and $\varphi_2$ fields are the ``Liouville fields,'' or more
properly, Toda fields, of the $W$-gravity sector.  The BRST operator
generalises the one given in [20] for the non-critical $W_3$ string [12,13].
Again $Q_B$ is graded, with $Q_0^2=Q_1^2=\{Q_0,Q_1\}=0$.  Our strategy for
obtaining the BRST operator consists of  writing down the most general possible
form for the operator  $F(\varphi_1,\beta,\gamma,T_M,W_M)$, which has conformal
weight 4 and ghost number 0, and imposing the conditions for nilpotence of
$Q_B$.  The condition $Q_0^2=0$ implies that the total central charge vanishes,
{\it i.e.}\
$$
12 \a^2 + 12 a^2+c_M-170=0\ ,\eqno(3.7)
$$
where $c_M$ is the central charge of the matter. We find that the remaining
nilpotency  conditions $\{Q_0,Q_1\}=Q_1^2=0$ imply
$$
2(\a-2 a)(\a-3 a)+1=0 \ ,\eqno(3.8)
$$
and they determine the coefficients of the terms in
$F(\varphi_1,\beta,\gamma,T_M,W_M)$ to be such that
$$
\eqalign{
Q_1=\oint dz\,\gamma &\Big[ (\df_1)^4 +  4\,\a\,
(\df_1)^2\,\del^2\varphi_1 +
  \ft14 \, ({\scriptstyle 130-c_M-8 \a^2 })\, \del^2\varphi_1\,\del^2\varphi_1
\cr\noalign{\vskip1pt}
  &+\ft16\, ({\scriptstyle -242+c_M+24 \a^2})\, \df_1\,\del^3\varphi_1 +
   \ft1{180 \a}\,({\scriptstyle -43092+252 c_M+2150 \a^2 +5 c_M\a^2+120\a^4})
\, \del^4\varphi_1 \cr\noalign{\vskip5pt}
&+ \ft{3}{5\a}\,({\scriptstyle -171+c_M+20\a^2})\, \df_1\,\del\beta\,\del\gamma
+ \ft{4}{5\a}\, ({\scriptstyle 513-3 c_M-40 \a^2 })\,
\del^2\varphi_1\,\beta\,\del\gamma \cr\noalign{\vskip5pt}
&+ \ft12\, ({\scriptstyle -196+c_M+16\a^2})\, \beta\,\del^3\gamma +
\ft1{15}({\scriptstyle 809-4 c_M-60 \a^2 })\,\del^2\beta\,\del\gamma
-  4\,T_M\,\beta\,\del\gamma \cr\noalign{\vskip5pt}
&- 2\,(\df_1)^2\,T_M + \ft{1}{5 \a}\,
({\scriptstyle 171-c_M-20\a^2})\, \df_1\,  \del T_M + \ft{2}{5\a} \,
({\scriptstyle -171+c_M+10\a^2})\, \del^2\varphi_1\,T_M \cr\noalign{\vskip5pt}
&+
\ft{2}{5}\,\ft{(-5051+36 c_M+420 \a^2)}{(22+c_M)}\, (T_M)^2 +
\ft15 \,\ft{(4091+215 c_M-c_M^2-340\a^2-20 c_M\a^2)}{(22+5
c_M)}\,\del^2 T_M \cr\noalign{\vskip5pt}
&+ ({\scriptstyle 240\a^2+17
c_M-2902 })\,\sqrt{\ft{(7 c_M+68)(4 c_M-2)
(c_M+24)}{75(22+5 c_M)(196-172 c_M+c_M^2)}} \, W_M
\Big]\ .\cr} \eqno(3.9)
$$

     It is convenient to express the background charges $\a$ and $a$, which are
related by (3.8), in terms of a parameter $t$:
$$
\a={2\over t}  +{3t\over2}\ ,\qquad
a={1\over t} +{t\over2}\ . \eqno(3.10)
$$
{}From (3.7), it then follows that
$$
c_M=86 -{60\over t^2} -30\, t^2\ .\eqno(3.11)
$$
Note that if we were to set $c_M=0$, in which case $T_M$ and $W_M$ would be
null fields that could be set to zero, (3.5) and (3.9) would reduce to the
critical \w24 BRST operator that we discussed in section 2.  Indeed, (3.11)
would then give $t^2=\ft53$ or $\ft65$.  The first of these implies that
$\a^2=\ft{243}{20}$, $a^2=\ft{121}{60}$, and (3.9) reduces to the BRST operator
given by (2.9).  The second solution, $t^2=\ft65$, gives $\a^2=\ft{361}{30}$,
$a^2=\ft{32}{15}$; this was also found in [15], and appears to be associated
with a critical string theory that is non-unitary in the multi-scalar case,
in that some of the longitudinal modes of spacetime-excited states have
negative
norms.  As a two-scalar theory, however, there is presumably no unitarity
problem.  The non-critical BRST operator that we have constructed here
has the property that it interpolates between the two critical \w24 BRST
operators found in [15].  The fact that there are two inequivalent solutions
of $c_M=0$ is related to the fact that the $B_2$ algebra is not simply
laced.  Both of the critical BRST operators are associated with the $W\!B_2$
algebra.

      In order to build a non-critical \w24 string theory, we need an explicit
realisation for the matter currents $T_M$ and $W_M$ appearing in (3.5) and
(3.9) above.  A two-scalar realisation of the $W\!B_2$ algebra was given in
[17], and takes the form
$$
\eqalign{
T_M&=T_{X_1}+ T_{X_2}
     =-\ft12(\del X_1)^2+ i\,({2\over t} -{3t\over2})\,\del^2 X_1
\,  -\ft12(\del X_2)^2+ i\,({1\over t} -{t\over2})\,
\del^2 X_2\ ,\cr\cr
W_M&=\lambda \,\Big[ g_1\, (\del X_1)^4 + g_2\, \del^2
X_1\,(\del X_1)^2 + g_3 \, (\del^2 X_1)^2 + g_4\, (\del^3 X_1)\, \del X_1
\cr   &\qquad\ \ +
g_5\, \del^4 X_1
g_6\, (\del X_1)^2\, T_{X_2} + g_7\, (\del X_1)\, (\del T_{X_2})
\cr   &\qquad\ \  +
g_8\, (\del^2 X_1)\,
T_{X_2} + g_9\, (T_{X_2})^2 + g_{10}\, \del^2 T_{X_2}\Big]\ ,\cr}
\eqno(3.12)
$$
where the coefficients $g_i$ are given by
$$
\eqalignno{
g_1&= \ft1{16}\, {\scriptstyle{t^2}\left( 3 - {t^2} \right) \,
       \left( -32 + 27\,{t^2} \right)} \ ,\cr
g_2&= -\ft{i}8\,{\scriptstyle t\,\left( -3 + {t^2} \right) \,
    \left( -4 + 3\,{t^2} \right) \,\left( -32 + 27\,{t^2} \right)} \ ,\cr
g_3&= \ft1{16}\,{\scriptstyle \left( -3 + {t^2} \right) \,
       \left( -152 + 336\,{t^2} - 246\,{t^4} + 63\,{t^6} \right)}
  \ ,\cr
g_4&= \ft14\, {\scriptstyle \left( -3 + {t^2} \right) \,
       \left( -60 + 144\,{t^2} - 115\,{t^4} + 30\,{t^6} \right) }
     \ ,\cr
g_5&= \ft{i}{48}\,{{\scriptstyle 1}\over {\scriptstyle t} }\, {\scriptstyle
\left( -3 + {t^2} \right) \,
       \left( -4 + 3\,{t^2} \right) \,
       \left( -60 + 144\,{t^2} - 115\,{t^4} + 30\,{t^6} \right) }\ ,\cr
g_6&= \ft34\, {\scriptstyle {t^2}\,\left( -68 + 113\,{t^2} - 41\,{t^4}
    \right)}\ ,&(3.13)\cr
g_7&= -\ft{i}2\, {\scriptstyle \left( -1 + t \right) \,t\,\left(1+ t \right) \,
    \left( 150 - 226\,{t^2} + 75\,{t^4} \right) }\ ,\cr
g_8&= -\ft{i}4\, {\scriptstyle t\,\left( -216 + 464\,{t^2} - 305\,{t^4} +
      69\,{t^6} \right)} \ ,\cr
g_9&= \ft14\, {\scriptstyle {t^2}\,\left( 3 - {t^2} \right) \,
       \left( -32 + 27\,{t^2} \right)} \ ,\cr
g_{10}&=\ft18 {\scriptstyle \left(240 - 724\,{t^2} + 865\, {t^4} - 465\,{t^6} +
90\,{t^8}\right)} \ ,\cr}
$$
and $\lambda$ is a normalisation constant given by
$$
\lambda^{-2}={{\scriptstyle 1}\over {\scriptstyle 8t^2}}
({\scriptstyle 3-t^2}) ({\scriptstyle -7 + 3 t^2})
({\scriptstyle -5 + 3 t^2}) ({\scriptstyle -2 + 3 t^2})
({\scriptstyle -5 + 4 t^2}) ({\scriptstyle -8 + 5 t^2})
({\scriptstyle -6 + 5 t^2}) ({\scriptstyle -6 + 7 t^2})
({\scriptstyle 150 -226t^2 + 75 t^4}) \ .\eqno(3.14)
$$
The realisation (3.12) generates the $W\!B_2$ algebra with central charge given
by (3.11). Substituting (3.12) into (3.9) gives the final result for $Q_1$ for
the  non-critical \w24 string.

     It should be emphasised that in the case of the critical \w24
string in section 2, the fact that there exists a closed $W\!B_2$ algebra at
the
quantum level did not appear to play an essential r\^ole in the construction of
the BRST operator.  Indeed, as we remarked earlier, the analogous construction
of a critical $W_{2,s}$ BRST operator can be carried out even for values of $s$
for which no closed algebra (at least not with the correct central
charge) exists [15].  For the non-critical BRST operator of this section, on
the
other hand, it is essential that the matter currents $T_M$ and $W_M$ do
generate
a closed algebra, namely $W\!B_2$,  at the quantum level.

\np
\noindent{\bf 4. Physical States in the Non-critical \w24 String}
\bigskip

     We have already remarked that the \w24 algebra can be obtained from
Hamiltonian reduction for the algebra $B_2$.  For convenience we shall begin
this section by setting up some notation.  We take the two simple roots of
the $B_2$  algebra to be
$$
e_1=(0,\,1)\,\qquad e_2=(1,\,-1)\ .\eqno(4.1)
$$
The associated fundamental weights $\lambda_i$, defined by ${2e_i\cdot
\lambda_j\over{(e_i\cdot e_i)}}=\delta_{ij}$ are
$$
\lambda_1=(\ft12,\,\ft12)\ ,\qquad \lambda_2=(1,\,0)\ .\eqno(4.2)
$$
Since $B_2$ is not simply-laced, we need also to introduce the simple
co-roots, defined by $e^\vee_i={2e_i\over{(e_i\cdot e_i)}}$, and the the
fundamental co-weights $\lambda^\vee_j$, defined by ${2e^\vee_i\cdot
\lambda^\vee_j\over{(e^\vee_i\cdot e^\vee_i)}}=\delta_{ij}$:
$$
\eqalignno{
e^\vee_1&=(0,\,2)\ ,\qquad e^\vee_2=(1,\, -1)\ , &(4.3)\cr
\lambda^\vee_1&=(1,\,1)\ ,\qquad \lambda^\vee_2=(1,\,0)\ .&(4.4)\cr}
$$
The Weyl vector and the Weyl co-vector are consequently
$$
\rho=(\ft32,\,\ft12)\ ,\qquad \rho^\vee=(2,\,1)\ .\eqno(4.5)
$$
In this language, the \w24 matter energy-momentum tensor in (3.12) becomes
$$
T_M=-\ft12 \del X\cdot\del X -i(t\,\rho-{{\scriptstyle 1}\over {\scriptstyle
t}}\,\rho^\vee)\cdot \del^2 X\ .\eqno(4.6)
$$
Note that since $\rho$ and $\rho^\vee$ are independent vectors, there is no
choice of the parameter $t$ for which the background charge vector in (4.6)
vanishes.  This contrasts with the situation for the $W_3$ algebra, where,
since $A_2$ is simply laced, there is no distinction between $\rho$ and
$\rho^\vee$.  A related difference is that here one can, for purely real or
purely imaginary $t$, only achieve matter central charges within certain
ranges.
Specifically, as may be seen from (3.11), if $t$ is real then we must have
$c_M\le 86-60\sqrt2 \approx 1.147$, whilst if $t$ is imaginary then $c_M\ge 86
+60\sqrt2 \approx 170.853$.  We note also that the energy-momentum tensor in
the gravity sector can be written as $T_L=T_{\varphi_1}+T_{\varphi_2}= -\ft12
\del\varphi\cdot\del\varphi -(t\,\rho+{{\scriptstyle 1}\over {\scriptstyle
t}}\,\rho^\vee)\cdot \del^2\varphi$.

     One can easily verify that there are four screening currents, {\it
i.e.}\ vertex operators that are total derivatives under the matter currents
(3.12):
$$\eqalignno{
S^+_j&=e^{-i\,\a_+\,e_j\cdot X}\ ,\qquad
S^-_j=e^{-i\,\a_-\,e^\vee_j\cdot X}\ ,&(4.7)\cr
\noalign{where}
\a_+&=t\ ,\qquad \a_-=-\, {{\scriptstyle 1}\over{\scriptstyle t}} \ .&(4.8)\cr}
$$
By following standard arguments, we find that vertex operators
$$
V^M_{r_i;s_i}=e^{i\,p^M\cdot X} \eqno(4.9)
$$
describing primaries of the \w24 minimal models have momenta given by
$$
p^M=\sum_{i=1}^2 \Big[(r_i-1)\a_+\,\lambda_i+(s_i-1)\a_-\,\lambda^\vee_i\Big]
\ .\eqno(4.10)
$$
We also define vertex operators
$$
V^L_{r_i;s_i}=e^{p^L\cdot \varphi} \eqno(4.11)
$$
for the Liouville sector, with momenta given by
$$
p^L=\sum_{i=1}^2 \Big[(r_i-1)\a_+\,\lambda_i- (s_i-1)\a_-\,\lambda^\vee_i\Big]
\ .\eqno(4.12)
$$

     We begin our discussion of physical states by looking at level $\ell=0$,
{\it i.e.}\ tachyons.  The form of these physical operators is
$$
U=c\, \del^2\gamma\, \del\gamma\, \gamma\, e^{p^L_1\,\varphi_1+p^L_2\,\varphi_2
+ i\, p^M_1\,X_1 + i\, p^M_2\,X_2}
\ .\eqno(4.13)
$$
For operators of this kind, where the $(b,c)$ dependence occurs just as an
overall factor of $c$, the grading of $Q_B$ implies that we may consider the
$Q_0$ and $Q_1$ parts of the BRST operator separately.  Acting with $Q_0$ gives
the mass-shell condition, which can be written as
$$
(\hat p^L_1)^2 + (\hat
p^L_2)^2 = (\hat p^M_1)^2 + (\hat p^M_2)^2\ ,\eqno(4.14)
$$
where the hatted momenta are shifted by the background charges:
$$
\hat p^L_1\equiv p^L_1+\a^L,\qquad \hat p^L_2\equiv p^L_1+a^L, \qquad
\hat p^M_1\equiv p^M_1+\a^M,\qquad \hat p^M_2\equiv p^L_1+a^M\ .\eqno(4.15)
$$
Here, $\a^L=\a$ and $a^L=a$ given by (3.10), and the background charges for the
matter sector, which can be read off from $T_M$ in (3.12), are
$$
\a^M={3t\over2}-{2\over t},\qquad a^M={t\over 2}-{1\over t}\ .\eqno(4.16)
$$
Acting with $Q_1$ on the operator (4.13) gives the further
physical-state condition
$$
(\hat p^L_1-\hat p^M_1)(\hat p^L_1+\hat p^M_1)(\hat p^L_1-\hat p^M_2)
(\hat p^L_1+\hat p^M_2)=0\ .\eqno(4.17)
$$

     There are eight classes of tachyons, corresponding to the four
different roots of (4.17) with, as a consequence of (4.14),
$\hat p^L_2=\pm \hat p^M_2$ (for each of the first two roots of (4.17)) or
$\hat p^L_2=\pm \hat p^M_1$ (for each of the remaining two roots) respectively.
The physical operators (4.13)  can conveniently be written in terms of the
vertex operators (4.9) and (4.11) in the eight cases:
$$
\eqalign{
t_1&=f\, V^L_{-r_1,-r_2;s_1,s_2}\ , \cr
t_3&=f\, V^L_{r_1,r_2;-s_1,-s_2}\ , \cr
t_5&=f\, V^L_{-r_1-2r_2,r_2;s_1+s_2,-s_2}\ , \cr
t_7&=f\, V^L_{r_1+2r_2,-r_2;-s_1-s_2,s_2}\ , \cr}
\qquad
\eqalign{
t_2&=f\, V^L_{r_1,-r_1-r_2;-s_1,2s_1+s_2}\ , \cr
t_4&=f\, V^L_{-r_1,r_1+r_2;s_1,-2s_1-s_2}\ , \cr
t_6&=f\, V^L_{r_1+2r_2,-r_1-r_2;-s_1-s_2,2s_1+s_2}\ , \cr
t_8&=f\, V^L_{-r_1-2r_2,r_1+r_2;s_1+s_2,-2s_1-s_2}\ , \cr}\eqno(4.18)
$$
where $f=c\,\del^2\gamma\,\del\gamma\,\gamma\, V^M_{r_1,r_2;s_1,s_2}$ in each
case.  Note that there is some redundancy in the parametrisation of the momenta
in (4.18), since the physical-state conditions (4.14) and (4.17) place two
conditions on the four momentum components in (4.13). Thus there are really
just two independent free parameters in the tachyon solutions.  Nonetheless, as
in [12] we find that the parametrisation (4.18) is a useful one.

     Turning now to higher-level physical states, we expect that there should
exist a ground-ring structure of physical operators at ghost number zero,
analogous to those that exist in the two-scalar Virasoro string [22],
two-scalar
$W_3$ string [23], and non-critical $W_3$ string [12].  Indeed, in [15] it was
found that there exists a physical operator at level $\ell=10$ and ghost number
$G=0$ in the multi-scalar (critical) $W_{2,4}$ string, which has zero momentum
in the effective spacetime.  This operator, $D$, is associated with the very
simple screening current $S\equiv\oint dw\, b(w) D = \beta\, e^{\ft29 \alpha
\varphi_1}$ [15], which satisfies $\{Q_B,\, S\}=\del D$.  This physical
operator
and screening current of course also exist in the special case of the
two-scalar
critical $W_{2,4}$ string.

     In the $W_3$ string, ground-ring operators exist in the
non-critical case [12] at the same level number, $\ell=6$, as ghost-number zero
discrete operators in the critical $W_3$ string [23].  Thus it is natural, by
analogy, to begin our search for ground-ring operators in the non-critical
$W_{2,4}$ string at level $\ell=10$.  They can be found by writing down the
most general structure at this level and ghost number, and then requiring that
$Q_B$ given by (3.4), (3.5), (3.9), (3.12) annihilate it.  In view of the
complexity of some of the calculations, it can be advantageous in practice to
begin by constructing the most general possible form for the discrete state
$D$, then acting with $\oint dw\, b(w)$ to obtain the associated screening
current $S$, and then solving the equation $\{Q_B,\, S\}=\del D$.  By using
this
procedure one has only to calculate the OPE of the BRST current with the
relatively-simple screening current $S$, rather than with the more complicated
physical operator $D$. At level $\ell=10$, we find that there are two discrete
physical operators in the non-critical $W_{2,4}$ string.  The first of these,
which we shall call $x$, is associated with the screening current $S_x$ given
by
$$
S_x=\beta\, V^L_{1112}\, V^M_{1112}\ .\eqno(4.19)
$$
This is the generalisation of the $\ell=10$ discrete operator of the critical
$W_{2,4}$ string that we mentioned above.  The second $\ell=10$ discrete
physical operator of the non-critical $W_{2,4}$ string, which we shall call
$y$, has a more complicated structure.  The associated screening current $S_y$
takes the form
$$
S_y=\big( \beta+\cdots\big) \, V^L_{2111}\, V^M_{2111}\ ,\eqno(4.20)
$$
where ``$\cdots$'' indicates 34 omitted terms, all of which have spin 4 and
ghost number $-1$.  The coefficients, which we have calculated, are
$t$-dependent.  We shall not present them here owing to the complexity of the
expression.

     By analogy with the non-critical $W_3$ string, where there are four
ground-ring operators for generic values of the matter central charge [12],
we might expect that there should be more ground-ring generators in our \w24
case, in addition to the two described above.  To find them, we note that a
feature of the ground-ring generators of the Virasoro and the $W_3$ strings
is that they can normal order with tachyonic physical operators to give
tachyonic physical operators again. For example, in the two-scalar Virasoro
string there are two branches of tachyons, corresponding to the two
solutions, $\hat p_L=\pm \hat p_M$, of the mass-shell condition $\hat
p_L^2=\hat p_M^2$.  Of the two $\ell=2$ ground-ring generators, one can
normal order with one of the tachyon branches to give a tachyon in the same
branch, whilst the other ground-ring generator can do the same for the
tachyons of the other branch.  In our $W_{2,4}$ case we have the eight
branches of tachyons $t_i$ listed in (4.18), corresponding to the eight
classes of solutions of (4.14) and (4.17).  It is easy to see that the $x$
operator associated with (4.19) can map $t_1$ and $t_2$ into themselves,
whilst $y$, associated with (4.20), can map $t_3$ and $t_7$ into themselves.
 By considering what other discrete operators might be able to map tachyons
in other branches, we find that the vertex operator $V^L_{1211}\,
V^M_{1211}$ has momentum that could map the tachyons $t_3$ and $t_4$, whilst
the vertex operator $V^L_{1121}\,  V^M_{1121}$ has momentum that could map
$t_5$.  From the spin of these vertex operators, it is easy to see that they
would have to be associated with $\ell=11$ states.  In fact we find that the
first, which we shall call $\tilde x$, does indeed exist, and has a
relatively simple structure, with its screening current $S_{\tilde x}$ given
by
$$
S_{\tilde x}=\Big (\beta\, \del\varphi_1 +({1\over 2 t}+{t\over 2})\, \del
\beta
\Big)\, V^L_{1211}\, V^M_{1211}\ .\eqno(4.21)
$$
The second, which we shall call $\tilde y$, is much more complicated, with
$S_{\tilde y}$ of the form
$$
S_{\tilde y}=\big (\beta\, \del\varphi_1 +\cdots
\big)\, V^L_{1121}\, V^M_{1121}\ .\eqno(4.22)
$$
We have not been able to carry out the computation in this case, but it seems
highly plausible that such a ground-ring generator exists.

     In the case of the non-critical $W_3$ string, it was shown in [21] that
whilst there are just four ground-ring operators for generic values of the
matter central charge [12], the number enlarges to six in the special case
$c_M=2$, which corresponds to vanishing background charge in the matter
sector.  In the case of \w24 on the other hand, we have already noted that
there is no choice of the parameter $t$ in (4.6) for which the background
charge totally vanishes in the matter sector.  It is not clear therefore
whether
one  should expect any additional ground-ring operators at any special values
of
$c_M$.

     In [12,21] the structure of the higher-level states with non-standard
ghost structure in the non-critical $W_3$ string is discussed. In
particular, the subset of physical states that exist in the special case of
$c_M=2$ can be grouped together into multiplets whose members are related by
transformations under the Weyl group of $SU(3)$. A similar discussion can be
given for the \w24 string, where in this case the Weyl group is that of the
underlying $B_2$ algebra.  It acts on the momenta of tachyonic vertex
operators according to the rule
$$
V^L_{r_i;-s_i}\, V^M_{r_i;s_i}\longrightarrow
V^L_{r_i;w\ast (-s_i)}\, V^M_{r_i;s_i}\ ,\eqno(4.23)
$$
with the Liouville momentum (4.12) transformed according to the rule
$$
\sum_{i=1}^2 \Big[(r_i-1)\a_+\,\lambda_i-(s_i-1)\a_-\,\lambda^\vee_i\Big]
\longrightarrow
\sum_{i=1}^2 \Big[(r_i-1)\a_+\,\lambda_i- s_i\,\a_-\,w\ast(\lambda^\vee_i)
 + \a_-\, \lambda^\vee_i \Big]\ .\eqno(4.24)
$$
Here, the Weyl transformations of the fundamental co-weights are dual to the
standard Weyl transformations of the simple co-roots, and are thus generated
by $w_1$ and $w_2$ defined by
$$
\eqalign{
w_1\ast(\lambda^\vee_1)&=2\lambda^\vee_2-\lambda^\vee_1,\qquad
w_1\ast(\lambda^\vee_2)=\lambda^\vee_2\ ,\cr
w_2\ast(\lambda^\vee_1)&=\lambda^\vee_1,\qquad\qquad\quad
w_2\ast(\lambda^\vee_2)=\lambda^\vee_1-\lambda^\vee_2\ ,\cr}\eqno(4.25)
$$
{}From (4.24), it follows that the action of the Weyl group on the
$(s_1,s_2)$ indices of $V^L_{r_i;-s_i}$ is given by
$$
w_1\ast(s_1,s_2)=(-s_1,2s_1+s_2),\qquad w_2\ast(s_1,s_2)=
(s_1+s_2,-s_2)\ .\eqno(4.26)
$$

     The elements $\{1,\, w_1,\, w_2,\, w_1\,  w_2,\,
w_2,\, w_1,\, w_2\, w_1\, w_2,\, w_1\, w_2\, w_1,\,  w_1\, w_2\, w_1\, w_2 \}$
generate the Weyl group of $B_2$.
Starting from the tachyons $t_3$ at level  $\ell=0$, the levels of the 8
Weyl-related states under the action (4.23) are $\{0,\, r_1 s_1,\,  r_2 s_2,\,
r_1 s_1 + r_1 s_2 + r_2 s_2,\, r_1 s_1 + 2 r_2 s_1 + r_2 s_2,\,
(r_1+2r_2)(s_1+s_2),\, (r_1+r_2)(2s_1+s_2),\, 2r_1s_1 + 2 r_2 s_1 + r_1 s_2  +
2r_2 s_2\}$.  The ghost numbers of the operators are $\{4,\, 3,\, 3,\,  2,\,
2,\, 1,\, 1,\, 0\}$.  The last case here corresponds to the ground-ring
operators.  From (4.26), it follows that the ground-ring operators have the
form $V^L_{r_i;s_i}\, V^M_{r_i;s_i}$.  The basic ground-ring generators
correspond to $(r_1,r_2,s_1,s_2)= (2,1,1,1)$, $(1,2,1,1)$, $(1,1,2,1)$ and
$(1,1,1,2)$, with levels $\ell=10$, 11, 11, 10 respectively.  In fact, these
correspond to the physical operators $y$, $\tilde x$, $\tilde y$ and $x$  that
we discussed previously.

     The eight classes of tachyon in (4.18) can be related to one another
under an action of the Weyl group that we have just been discussing.  First,
we define the Weyl group action on the $r_i$ indices of a vertex operator, in
a similar fashion to the action on $s_i$ indices described above:
$$
w_1\ast(r_1,r_2)=(-r_1,2r_1+r_2),\qquad w_2\ast(r_1,r_2)=
(r_1+r_2,-r_2)\ .\eqno(4.27)
$$
Defining $(W,\, W')\, V^L_{r_i;s_i}\equiv V^L_{W\ast (r_i); W'\ast (s_i)}$ as
the action of Weyl-group elements $W=w_{j_1}w_{j_2}\cdots$ and
$W'=w_{k_1}w_{k_2}\cdots$, acting on the Liouville vertex operators, we find
that we may write the tachyons in (4.18) as follows:
$$
\eqalign{
t_1&=(w_1 w_2 w_1 w_2, \, w_1 w_2 w_1 w_2)\, t_3\ ,\cr
t_3&=(1,\, 1) \, t_3\ ,\cr
t_5&=(w_1 w_2 w_1 , \, w_1 w_2 w_1 )\, t_3\ ,\cr
t_7&=(w_2,\, w_2) \, t_3\ ,\cr}
\qquad
\eqalign{
t_2&=( w_2 w_1 w_2, \, w_2 w_1 w_2)\, t_3\ ,\cr
t_4&=(w_1,\, w_1) \, t_3\ ,\cr
t_6&=(w_2 w_1 , \, w_2 w_1 )\, t_3\ ,\cr
t_8&=(w_1 w_2,\,w_1 w_2) \, t_3 \ . \cr}\eqno(4.28)
$$

     We have looked also at the physical states at level $\ell=1$, and find the
following physical operators
$$
\eqalignno{
c\, \del\gamma\, \gamma\, V^L_{-2-r,1,s,1}\, V^M_{r,1,s,1}& \quad\sim\quad
(1,\, w_2)\, t_5\cr
c\, \del\gamma\, \gamma\, V^L_{r,1,-1-s,1}\, V^M_{r,1,s,1}& \quad\sim\quad
(1,\, w_2)\, t_3\cr
c\, \del\gamma\, \gamma\, V^L_{2+r,-1-r,-s,1+2s}\, V^M_{r,1,s,1}&
\quad\sim\quad
(w_1,\,w_1w_2)\, t_5\cr
c\, \del\gamma\, \gamma\, V^L_{-r,1+r,1+s,-1-2s}\, V^M_{r,1,s,1}&
\quad\sim\quad
(w_1,\, w_1w_2)\, t_3\cr
\cr
c\, \del\gamma\, \gamma\, V^L_{-r,1,s-1,1}\, V^M_{r,1-r,s,1-2s}& \quad\sim\quad
(1,\, w_2)\, t_4\cr
c\, \del\gamma\, \gamma\, V^L_{r-2,1,-s,1}\, V^M_{r,1-r,s,1-2s}& \quad\sim\quad
(1,\, w_2)\, t_8\cr
c\, \del\gamma\, \gamma\, V^L_{r,1-r,1-s,2s-1}\, V^M_{r,1-r,s,1-2s}&
\quad\sim\quad
(w_1,\, w_1w_2)\, t_4\cr
c\, \del\gamma\, \gamma\, V^L_{2-r,r-1,s,1-2s}\, V^M_{r,1-r,s,1-2s}&
\quad\sim\quad
(w_1,\, w_1w_2)\, t_8 &(4.29)\cr
\cr
c\, \del\gamma\, \gamma\, V^L_{2r-1,1-r,1-s,s}\, V^M_{1-2r,r,1-s,s}&
\quad\sim\quad
(w_2,\, w_2w_1)\, t_6\cr
c\, \del\gamma\, \gamma\, V^L_{1-2r,r,s-1,2-s}\, V^M_{1-2r,r,1-s,s}&
\quad\sim\quad
(w_2,\, w_2w_1)\, t_7\cr
\cr
c\, \del\gamma\, \gamma\, V^L_{2-2r,r,s-\ft12,1-s}\, V^M_{2-2r,r,\ft12-s,s}&
\quad\sim\quad (w_2,\, w_2w_1)\, t_7\cr
c\, \del\gamma\, \gamma\, V^L_{2r-2,2-r,\ft12-s,s}\, V^M_{2-2r,r,\ft12-s,s}&
\quad\sim\quad (w_2,\, w_2w_1)\, t_6\cr
\cr
c\, \del\gamma\, \gamma\, V^L_{-3,2,2,-1}\, V^M_{1,1,1,1}&
\quad\sim\quad (w_2,\, w_2 w_1)\, t_2 \cr
c\, \del\gamma\, \gamma\, V^L_{3,-1,-2,3}\, V^M_{1,1,1,1}&
\quad\sim\quad (w_2,\, w_2 w_1)\, t_3 \cr
}
$$
As in the case of the tachyons,
there is redundancy in the parametrisation of the momentum (in each case, $r$
and $s$ occur in a fixed combination, so there is really only one continuous
parameter here), but nonetheless the parametrisation is a convenient one.
In addition, there are three
more physical operators of the form $\Big(c\, \del\gamma\, \gamma + \lambda \,
\del^2\gamma\, \del\gamma\, \gamma\Big)\, V^L\, V^M$, with discrete
momenta, where the vertex operators are
$$
\eqalign{
V^L_{1,-1-r,1,s}\,V^M_{1,r,1,s}&\quad\sim\quad (1,\, w_1)\, t_2   \cr
V^L_{1,r,1,-2-s}\,V^M_{1,r,1,s}&\quad\sim\quad  (1,\, w_1)\, t_3   \cr
V^L_{1,-1,1,-1}\,V^M_{1001}&\quad\sim\quad (1,\, w_1)\,t_6    \cr}\eqno(4.30)
$$
In both (4.29) and (4.30), the expression to the right of the $\sim$ symbol
indicates how the momentum of the state may be related to that of a tachyon by
a Weyl-group transformation, according to the rule $(W,\, W')\,
V^L_{r_i;s_i}\equiv  V^L_{W\ast (r_i); W'\ast (s_i)}$.

\bigskip\bigskip
\noindent{\bf 5. The Cohomology of the Critical $W_{2,4}$ String.}
\bigskip

     One would like to have a systematic procedure for determining the spectra
of physical states in the various $W$-string theories.  For the non-critical
$W_{2,4}$ string of section 4, this would be a very complicated problem, which
we shall not attempt to solve here.  Methods analogous to those used in [24]
for
the two-scalar Virasoro string, and in [21] for the non-critical $W_3$ string,
could presumably be used.  For the critical $W_{2,4}$ string on the  other
hand, we can use a method that was recently discovered for solving the
cohomology of the one-scalar Virasoro string and the two-scalar critical $W_3$
string [11].  The basic idea is similar to that of the ground-ring construction
for the two-scalar Virasoro string [22], in that higher-level physical
operators
are obtained by acting with powers of certain generators (which are themselves
physical operators).  However, there are important differences from the case of
the two-scalar string.  First of all, in our case the generators will be
physical operators with non-zero ghost numbers; these are the natural objects
to use in a theory where the spectrum of physical states fans out over a wider
and wider range of ghost numbers as one goes to higher and higher levels.
Secondly, and most crucially, the generating operators will have inverses.  A
consequence of this is that, when powers of the generators are normal ordered
with any physical operator, the result is guaranteed to be a BRST non-trivial
physical operator.  Since the generators carry momentum, this means that {\it
all} physical operators of the theory can be mapped into physical operators in
a certain fundamental cell in the momentum plane, where an exhaustive
construction of all possible physical operators can be carried out.  Having
done this, acting with all possible powers of the generators will yield the
entire cohomology of physical states [11].

     We shall begin our discussion with the two-scalar critical \w24 string
described in section 2.  Many physical states for this theory were found in
[15]; a characteristic feature is that the momenta in the vertex operators
$e^{p_1\varphi_1 + p_2\varphi_2}$ of physical states are always quantised:
$$
(p_1,\, p_2)=(\ft1{27}\alpha\, k_1,\, \ft1{11} a\, k_2)\ ,\eqno(5.1)
$$
where $k_1$ and $k_2$ are integers.  For this reason, we find it convenient to
characterise the momenta of the physical states by $(k_1,\, k_2)$, and we shall
commonly call these the momenta.  The mass-shell condition for a physical
state at level $\ell$ therefore implies the relation
$$
(k_1+27)^2 + (k_2+11)^2 =10(12\ell+1)\ .\eqno(5.2)
$$
A necessary condition for any physical operator is that its momentum should
satisfy (5.2) for integer $k_1$, $k_2$ and $\ell$.  Thus candidates for the two
invertible physical operators that could generate the cohomology must have
momenta $(k_1,\, k_2)$ for which $(-k_1,\, -k_2)$ also satisfies (5.2).  They
must also be able to normal order with all other physical operators.  It is
not hard to see that the simplest two possibilities are to have operators,
which we shall call $X$ and $Y$, with momenta $(30,\, 0)$ and $(0,\, 30)$
respectively.  From (5.2), it follows that $X$ has level $\ell=28$,
and $Y$ has level $\ell=20$.  Their inverses, $X^{-1}$ and $Y^{-1}$,
have momenta $(-30,\, 0)$ and $(0,\,-30)$, and occur at levels $\ell=1$ and
$\ell=9$ respectively.

     Despite its high level number, the $X$ operator at $\ell=28$ is easily
constructed.  Its associated screening current, $S_X=\oint dw\, b(w) X$, takes
the simple form
$$
S_X=\del^3\beta\, \del^2\beta\, \del\beta\, \beta\, e^{\ft{10}9\alpha
\varphi_1}
\ . \eqno(5.3)
$$
It is quite straightforward now to verify that $[Q_B,S_X]=\del X$, thus
establishing that $X$ is a physical operator.   The form of $X$ itself is quite
complicated, and we shall not present it here.  We just remark that, as may be
seen from (5.3), $X$ has ghost number $G=-3$.   The inverse of the $X$ operator
is the very simple level 1 physical operator, with ghost number $G=3$;
$$
X^{-1}=\Big(c\, \del\gamma\, \gamma -\ft{28}{15}\del^2\gamma\, \del\gamma\,
\gamma\Big)\, e^{-\ft{10}9\alpha\varphi_1} .\eqno(5.4)
$$
It is a straightforward matter to verify that indeed the normal-ordered product
of $X^{-1}$ and $X$ gives the identity (up to an overall non-vanishing constant
factor, whose precise value is unimportant in the subsequent argument).
Another way of obtaining the $X$ operator is by noting that there exists a
$G=0$ physical operator $x$ at $\ell=10$, with momentum $(6,\, 0)$, with
corresponding screening current $S_x=\beta\, e^{\ft29\alpha\varphi_1}$ [15].
By
acting appropriately on $x$ with four of these screening currents (and a ghost
booster $[Q_B,\varphi_1]$), one can construct the $\ell=28$ operator $X$ that
we have described above.  The detailed calculation would be quite involved,
with care being needed to handle the multiple contour integerals properly, and
in practice the direct verification that $[Q_B,S_X]=\del X$ is much simpler.

     The $Y$ operator, and its associated screening current $S_Y$, is much more
complicated, and in fact we have not yet managed to compute it.  The reason for
this is that its ghost number is considerably higher (in fact $G=-1$) than the
lowest possible one at level $\ell=20$, and also, unlike the screening current
(5.3) for the $X$ operator, $S_Y$ will involve the $(b,\, c)$ ghosts.  Thus
there are a very large number of terms in the expression for $Y$.  Its inverse,
on the other hand, at level $\ell=9$ with $G=1$, is quite simple and was in
fact
already constructed  in [15].  It takes the form
$$
Y^{-1}= c\, U\, e^{-\ft{30}{11}a\varphi_2}\ ,\eqno(5.5)
$$
where $U$ is the spin-3 current
$$
\eqalign{
U&=\ft53\, (\del\varphi_1)^3 +5\alpha
\,\del^2\varphi_1\, \del\varphi_1 +\ft{25}4\, \del^3\varphi_1 + 20\,
\del\varphi_1\, \beta\, \del\gamma\cr
&\qquad+12 \, \del\varphi_1\, \del\beta\, \gamma + 12\, \del^2\varphi_1\,
\beta\, \gamma + 5\alpha \, \del\beta\, \del \gamma + 3\alpha
\, \del^2\beta\, \gamma\ .\cr}\eqno(5.6)
$$
Interestingly enough, as was shown in [16], the spin-3 current $U$ and the
energy-momentum tensor $T=T_{\varphi_1}+T_{\gamma,\beta}$ form the spin-3 and
spin-2 currents of a realisation of the $W_3$ algebra at $c=\ft45$.  For now,
we
shall proceed by  assuming that the $\ell=20$ physical operator $Y$ exists, and
that its  normal-ordered product with $Y^{-1}$ is a non-zero constant, which
may
be  taken to be unity by an appropriate rescaling.

      It was shown in [11] that if an invertible physical operator is normal
ordered with any BRST non-trivial physical operator $V$, then the result is
itself a BRST non-trivial physical operator.  This is because, modulo
BRST-trivial terms that do not affect the argument, if one normal orders,
for example, the invertible operator $X$ with $V$, then one can recover $V$
by a further normal ordering with $X^{-1}$:  {\it i.e.}\  $V\approx (X^{-1}\,
(X\, V))$.   Thus $(X\, V)$ itself must be BRST non-trivial, since if $P$ is
BRST-trivial, then so is $(X^{-1}\,P)$.

     As a consequence of the above argument, it follows that we may take any
physical operator in the critical two-scalar \w24 string, and map it into a
BRST non-trivial physical operator whose momentum $(k_1,\, k_2)$ lies in a
fundamental unit cell, of size 30 by 30, by acting with appropriate powers
(positive or negative) of the invertible operators $X$ and $Y$.  We find it
convenient to choose this fundamental unit cell to be the one defined by
$$
-32\le k_1\le -3,\qquad\qquad -25\le k_2\le 4\ .\eqno(5.7)
$$
If we now solve explicitly for all physical operators whose momenta lie in
this unit cell, it follows that the complete cohomology of physical operators
for the theory is then obtained by acting on this set of fundamental
physical operators with $X^m\, Y^n$ for all integer $m$ and $n$.

     The momentum $(k_1,\, k_2)$ of any physical operator must satisfy (5.2)
for integer $k_1$, $k_2$ and level number $\ell$.  It is an elementary
exercise to enumerate all the possible lattice points in the unit cell (5.7)
that satisfy this necessary condition for physical operators.  It turns out
that 36 of the 900 points fulfil this requirement, with level numbers lying
in the range $0\le\ell\le5$.  In fact our choice (5.7) for the fundamental
unit cell was motivated by the desideratum that all the candidate solutions
within it have  low level numbers.  Of the 36 candidates, 28 correspond to
physical  operators that were already found in [15]; these
generalise  to continuous-momentum operators the multi-scalar critical \w24
string.  They comprise eight tachyons $t_i$ at ghost number $G=4$, level
$\ell=0$; ten operators $u_i$ at $G=3$, with levels $\ell=1$ and 2; and ten
operators $v_i$ at $G=2$, with levels $\ell=3$, 4 and 5.  In addition, we find
that there are a further four operators $d_i$ at $\ell=1$, which would
generalise to physical operators with discrete spacetime momenta in the
multi-scalar case.  (We commonly refer to these as discrete operators, even in
the two-scalar case.)  One of these is the operator $X^{-1}$ itself, given in
(5.4).  Thus we have a total of 32 physical operators in the fundamental unit
cell.  In detail, they comprise eight $G=4$ operators at $\ell=0$:
$$
\eqalign{
t_1&= c\, \del^2\gamma\, \del\gamma\, \gamma\,
e^{-\ft{30}{27}\alpha\varphi_1 -\ft{12}{11}a\varphi_2}\ ,\cr
t_3&= c\, \del^2\gamma\, \del\gamma\, \gamma\,
e^{-\ft{24}{27}\alpha\varphi_1 -\ft{12}{11}a\varphi_2}\ ,\cr
t_5&= c\, \del^2\gamma\, \del\gamma\, \gamma\,
e^{-\ft{28}{27}\alpha\varphi_1 -\ft{14}{11}a\varphi_2}\ ,\cr
t_7&= c\, \del^2\gamma\,\del\gamma\, \gamma\,
e^{-\ft{26}{27}\alpha\varphi_1-\ft{14}{11}a\varphi_2}\ ,\cr }\qquad
\eqalign{
t_2&= c\, \del^2\gamma\, \del\gamma\, \gamma\,
e^{-\ft{30}{27}\alpha\varphi_1 -\ft{10}{11}a\varphi_2}\ ,\cr
t_4&= c\, \del^2\gamma\, \del\gamma\, \gamma\,
e^{-\ft{24}{27}\alpha\varphi_1 -\ft{10}{11}a\varphi_2}\ ,\cr
t_6&= c\, \del^2\gamma\, \del\gamma\, \gamma\,
e^{-\ft{28}{27}\alpha\varphi_1 -\ft{8}{11}a\varphi_2}\ ,\cr
t_8&= c\, \del^2\gamma\,\del\gamma\, \gamma\,
e^{-\ft{26}{27}\alpha\varphi_1-\ft{8}{11}a\varphi_2}\ ,\cr}
\eqno(5.8)
$$
six $G=3$ operators at $\ell=1$:
$$
\eqalign{
u_1&= c\,  \del\gamma\, \gamma\,
e^{-\ft{20}{27}\alpha\varphi_1 -\ft{20}{11}a\varphi_2}\ ,\cr
u_3&= c\,  \del\gamma\, \gamma\,
e^{-\ft{18}{27}\alpha\varphi_1 -\ft{18}{11}a\varphi_2}\ ,\cr
u_5&= c\,  \del\gamma\, \gamma\,
e^{-\ft{16}{27}\alpha\varphi_1 -\ft{14}{11}a\varphi_2}\ ,\cr }\qquad
\eqalign{
u_2&= c\, \del\gamma\, \gamma\,
e^{-\ft{20}{27}\alpha\varphi_1 -\ft{2}{11}a\varphi_2}\ ,\cr
u_4&= c\,  \del\gamma\, \gamma\,
e^{-\ft{18}{27}\alpha\varphi_1 -\ft{4}{11}a\varphi_2}\ ,\cr
u_6&= c\,  \del\gamma\, \gamma\,
e^{-\ft{16}{27}\alpha\varphi_1 -\ft{8}{11}a\varphi_2}\ ,\cr}
\eqno(5.9a)
$$
together with four more $G=3$ operators at $\ell=2$:
$$
\eqalign{
u_7&=c\, \Big( \del\varphi_1\, \del\gamma\, \gamma\, -\ft2{15}\alpha\,
\del^2\gamma\, \gamma\Big)\,
e^{-\ft{18}{27}\alpha\varphi_1 -\ft{24}{11}a\varphi_2}\ ,\cr
u_8&=c\, \Big( \del\varphi_1\, \del\gamma\, \gamma\, -\ft2{15}\alpha\,
\del^2\gamma\, \gamma\Big)\,
e^{-\ft{18}{27}\alpha\varphi_1 +\ft{2}{11}a\varphi_2}\ ,\cr
u_9&=c\, \Big( \del\varphi_1\, \del\gamma\, \gamma\, -\ft4{27}\alpha\,
\del^2\gamma\, \gamma\Big)\,
e^{-\ft{14}{27}\alpha\varphi_1 -\ft{20}{11}a\varphi_2}\ ,\cr
u_{10}&=c\, \Big( \del\varphi_1\, \del\gamma\, \gamma\, -\ft4{27}\alpha\,
\del^2\gamma\, \gamma\Big)\,
e^{-\ft{14}{27}\alpha\varphi_1 -\ft{2}{11}a\varphi_2}\ .\cr}\eqno(5.9b)
$$
At $G=2$ there are four operators with $\ell=3$:
$$
\eqalign{
v_1&=c\,\gamma\, e^{-\ft{10}{27}\alpha\varphi_1 -\ft{20}{11}a\varphi_2}\ ,\cr
v_3&=c\,\gamma\, e^{-\ft{8}{27}\alpha\varphi_1 -\ft{14}{11}a\varphi_2}\ ,\cr}
\qquad
\eqalign{
v_2&=c\,\gamma\, e^{-\ft{10}{27}\alpha\varphi_1 -\ft{2}{11}a\varphi_2}\ ,\cr
v_4&=c\,\gamma\, e^{-\ft{8}{27}\alpha\varphi_1 -\ft{8}{11}a\varphi_2}\ ,\cr}
\eqno(5.10a)
$$
two more at $\ell=4$:
$$
\eqalign{
v_5&=c\, \Big( \del\varphi_1\, \gamma\, -\ft8{27}\alpha\,
\del\gamma\Big)\,
e^{-\ft{6}{27}\alpha\varphi_1 -\ft{18}{11}a\varphi_2}\ ,\cr
v_6&=c\, \Big( \del\varphi_1\, \gamma\, -\ft8{27}\alpha\,
\del\gamma\Big)\,
e^{-\ft{6}{27}\alpha\varphi_1 -\ft{4}{11}a\varphi_2}\ ,\cr}\eqno(5.10b)
$$
and four further operators at $\ell=5$:
$$
\eqalign{
v_7&=c\, \Big(\beta\,\del\gamma\,\gamma -\ft{35}{108}\alpha\, \del\varphi_1\,
\del\gamma +\ft5{36}\alpha\, \del^2\varphi_1\, \gamma +\ft58\,
(\del\varphi_1)^2 \,\gamma \Big)\,
e^{-\ft{6}{27}\alpha\varphi_1 -\ft{24}{11}a\varphi_2}\ ,\cr
v_8&=c\, \Big(\beta\,\del\gamma\,\gamma -\ft{35}{108}\alpha\, \del\varphi_1\,
\del\gamma +\ft5{36}\alpha\, \del^2\varphi_1\, \gamma +\ft58\,
(\del\varphi_1)^2 \,\gamma \Big)\,
e^{-\ft{6}{27}\alpha\varphi_1 +\ft{2}{11}a\varphi_2}\ ,\cr
v_9&=c\, \Big(\beta\,\del\gamma\,\gamma -\ft{4}{9}\alpha\, \del\varphi_1\,
\del\gamma +\ft1{18}\alpha\, \del^2\varphi_1\, \gamma +\ft34\,
(\del\varphi_1)^2 \,\gamma  +\ft3{10} \del^2\gamma\Big)\,
e^{-\ft{4}{27}\alpha\varphi_1 -\ft{20}{11}a\varphi_2}\ ,\cr
v_{10}&=c\, \Big(\beta\,\del\gamma\,\gamma -\ft{4}{9}\alpha\, \del\varphi_1\,
\del\gamma +\ft1{18}\alpha\, \del^2\varphi_1\, \gamma +\ft34\,
(\del\varphi_1)^2 \,\gamma  +\ft3{10} \del^2\gamma\Big)\,
e^{-\ft{4}{27}\alpha\varphi_1 -\ft{2}{11}a\varphi_2}\ .\cr
}\eqno(5.10c)
$$
The four additional discrete operators in the
fundamental unit cell, at $\ell=1$, are as follows:
$$
\eqalign{
d_1&=\Big( c\,\del\gamma\, \gamma -\ft{28}{15}\, \del^2\gamma\, \del\gamma\,
\gamma \Big)\, e^{-\ft{30}{27}\alpha\varphi_1}\ ,\cr
d_2&=\Big( c\,\del\gamma\, \gamma -\ft{4}{15}\, \del^2\gamma\, \del\gamma\,
\gamma \Big)\, e^{-\ft{24}{27}\alpha\varphi_1}\ ,\cr
d_3&=c\, \Big( \del^4\gamma\, \del^2\gamma\, \del\gamma\, \gamma -\ft{10}9 \,
\del\varphi_1\, \del^3\gamma\, \del^2\gamma\, \del\gamma\, \gamma \Big)
e^{-\ft{30}{27}\alpha\varphi_1-\ft{22}{11}a \varphi_2}\ ,\cr
d_4&=c\, \Big( \del^4\gamma\, \del^2\gamma\, \del\gamma\, \gamma -\ft{8}9 \,
\del\varphi_1\, \del^3\gamma\, \del^2\gamma\, \del\gamma\, \gamma \Big)
e^{-\ft{24}{27}\alpha\varphi_1-\ft{22}{11}a \varphi_2}\ .\cr}\eqno(5.11)
$$
The physical operators $d_1$ and $d_2$ have $G=3$, whilst $d_3$ and $d_4$ have
$G=5$.  The operator $d_1$ is the same as $X^{-1}$.  The remaining four
candidate momenta in the fundamental cell (5.7) do not correspond to any
physical operators.

      The operators that we have listed above are all {\it prime}
operators, {\it i.e.}\ they are the physical operators, for given momentum,
with
the lowest possible ghost number.  As has been discussed extensively in
[22,23,9,10,11], each such operator is associated with a quartet of physical
operators, whose remaining three members are obtained by normal-ordering the
prime operator with $a_{\varphi_1}$, $a_{\varphi_2}$ or
$a_{\varphi_1}a_{\varphi_2}$, where $\del a_{\varphi_i}\equiv [Q_B,\,
\del\varphi_i]$.  Thus starting from a prime state at ghost number $G$, the
quartet of operators has ghost numbers $\{G,\, G+1,\, G+1,\, G+2\}$.  We shall
just focus our discussion on the prime operators, it being understood that
there
is a quartet associated with each prime operator.

     The complete cohomology of prime operators for the two-scalar critical
\w24 string is then given by
$$
\eqalign{
X^m\,Y^n\, t_i&:\qquad\qquad G=4-3m-n\ ,\cr
X^m\,Y^n\, u_i&:\qquad\qquad G=3-3m-n\ ,\cr
X^m\,Y^n\, v_i&:\qquad\qquad G=2-3m-n\ ,\cr
X^m\,Y^n\, d_1&:\qquad\qquad G=3-3m-n\ ,\cr
X^m\,Y^n\, d_2&:\qquad\qquad G=3-3m-n\ ,\cr
X^m\,Y^n\, d_3&:\qquad\qquad G=5-3m-n\ ,\cr
X^m\,Y^n\, d_4&:\qquad\qquad G=5-3m-n\ ,\cr}\eqno(5.12)
$$
In all cases, $m$ and
$n$ can be arbitrary integers.  The momenta $(k_1',\, k_2')$ of the resulting
physical operators are given by $(k_1',\, k_2')=(k_1+30m,\, k_2+30n)$, where
$(k_1,\, k_2)$ is the momentum of the original physical operator in the
fundamental unit cell.  Their level numbers are obtained by substituting
$(k_1',\, k_2')$ into (5.2).

     It was implicit in the above construction that the normal-ordered products
of $X$ and $Y$ with any physical operator are well-defined, in other words
that the operator products give integer-degree poles.   It is easy to check
that
this is indeed the case.  Recalling that the OPE of vertex operators with
momenta $p$ and $p'$ gives a factor $(z-w)^{-p\cdot p'}$, we see that if $X$ is
normal ordered with a physical operator whose momentum is given by $(k_1,\,
k_2)$, then the vertex operators will give a pole of degree $\ft12 k_1$.
Similarly, if $Y$ is normal ordered with the physical operator the vertex
operators will give a pole of degree $\ft12 k_2$.  One can easily see from the
mass-shell condition (5.2), which is a necessary condition that must be
satisfied by any physical operator, that $k_1$ and $k_2$ are always even
integers.  Thus we see that normal-ordering $X$ or $Y$ with any physical
operator will always give an integer-degree pole, and so the product is
well-defined.

     In view of the fact that we have only been able to conjecture the
existence
of the invertible operator $Y$, because of the complexity of the calculation
necessary for proving its existence, it is worthwhile to carry out some
consistency checks in order to verify the plausibility of the conjecture.  One
way to do this is to consider some of the physical operators that would be
generated by acting with $Y$ on the fundamental states.  It is easy to see from
(5.2) that if one acts with $Y$ on a fundamental operator with momentum
$(k_1,\,
k_2)$ at level $\ell$, one obtains a new physical operator with momentum
$(k_1,\,
k_2+30)$ at level $\ell'=\ell+13+\ft12 k_2$.  From this, we see that a simple
example is to consider $(Y\, u_7)$, where $u_7$ is the fundamental
$\ell=2$, $G=3$ operator given in (5.9$b$).  This would yield a physical
operator with momentum $(-18,\, 6)$ at level $\ell'=3$ and ghost number $G=2$.
In fact just such a physical operator exists, and was found in [15]:
$$
V=\Big(c\,\gamma + \ft43 \del^2\gamma\, \gamma -\ft{16}9 \alpha\,
\del\varphi_1\, \del \gamma\, \gamma-\ft8{11}a \,\del \varphi_2\, \del \gamma\,
\gamma -\ft85 b\, c\, \del\gamma\, \gamma \Big) e^{-\ft23 \alpha \varphi_1
+\ft6{11} a\varphi_2}\ .\eqno(5.13)
$$
The existence of this physical operator thus provides supporting evidence for
the existence of the $Y$ operator.  We have also checked some other examples of
a similar kind.  In particular we have verified that $(Y\, d_3)$ at $\ell=3$
and $G=4$; $(Y\, u_1)$ at $\ell=4$ and $G=2$; and $(Y\, v_1)$ and $(Y\, v_7)$
at
$\ell=6$ and $G=1$ all indeed correspond to actual physical operators.

     We are now in a position to discuss how the Weyl group of $B_2$ acts on
the physical states of the two-scalar critical \w24 string.  We first define a
momentum $(\hat k_1,\, \hat k_2)$ that is shifted by the background charges;
$(\hat k_1,\, \hat k_2)\equiv (k_1+27,\, k_2 +11)$, in terms of which the
mass-shell condition becomes
$$
\hat k_1^2 + \hat k_2^2 =10(12\ell+1)\ .\eqno(5.14)
$$
This is clearly invariant under the $B_2$ Weyl group, generated by
$$
\eqalign{
S_1&:\qquad (\hat k_1,\, \hat k_2)\longrightarrow (\hat k_1,\, -\hat k_2)\ ,\cr
S_2&:\qquad (\hat k_1,\, \hat k_2)\longrightarrow (\hat k_2,\, \hat k_1)\ .\cr}
\eqno(5.15)
$$
Of course equation (5.14) is in fact invariant under $O(2)$, but we are
interested in its $B_2$ Weyl subgroup since this also acts covariantly on the
other constraints arising from the physical-state conditions.  For example, in
the case of tachyons $t$ the physical-state condition $[Q_0,\, t]=0$ gives the
mass-shell condition (5.14) (with $\ell=0$), and the remaining physical-state
condition $[Q_1,\, t]=0$ implies the vanishing of the quartic polynomial
$$
W=(\hat k_1^2-1)(\hat k_1^2-9)\ , \eqno(5.16)
$$
which one can easily see is invariant under the Weyl group (after using the
mass-shell condition if necessary).  Thus, the set of eight tachyonic physical
states (5.8) are mapped into each other under the action of the Weyl group, as
may easily be verified using (5.15).  For higher-level physical states the
action of the Weyl group is a little more complicated, and in fact although it
still preserves the level number $\ell$, it now maps between prime physical
states at a set of different ghost numbers $G$ [11].  Its action is most easily
understood by noting that if one takes any integer solution of the mass-shell
equation (5.14) that actually corresponds to a physical state (recall that, for
example, in the fundamental cell (5.7), 32 of the 36 integer solutions
correspond to physical states), then after acting with any of the Weyl group
elements generated by (5.15) one arrives at a momentum that corresponds to
another physical state.  This can be verified by showing that the resulting
momentum is one that can be obtained by acting with some powers of $X$ and $Y$
on one of the fundamental operators given in (5.8)--(5.11).  Thus we see that
the Weyl group maps prime physical states into other prime physical states at
the same level.  As discussed in [11], the ghost numbers of these states will
be
unequal, except in the special case of the tachyons.

     So far in this section, we have concentrated on constructing the
cohomology of the two-scalar critical \w24 string.  We may also extend the
discussion to the multi-scalar critical \w24 string.  We recall that the
multi-scalar critical \w24 string [15] is obtained by replacing the
energy-momentum tensor $T_{\varphi_2}$ in (2.2) by
$$
T^{\rm eff}=-\ft12\, \del X^\mu\, \del X^\nu\, \eta_{\mu\nu} -i\, a_\mu\,
\del^2 X^\mu\ ,\eqno(5.17)
$$
where the background-charge vector $a_\mu$ is chosen so that $T^{\rm eff}$ has
the same central charge, namely $c^{\rm eff}=\ft{126}5$, as did $T_{\varphi_2}$
given by (2.5).

     As described in [11] for the case of $W_3$ strings, one may derive the
multi-scalar cohomology by considering the subset of two-scalar physical
operators that can be generalised to the multi-scalar case.  To be
generalisable, two-scalar physical operators must fall into one of the
following
three categories:

\medskip

\item{1)} If there is a pair of two-scalar prime physical operators with
momenta $(k_1,\, k_2)$ and $(k_1,\, -22 -k_2)$, both at the same ghost
number. This pair, which have the same conformal dimension
$\Delta=-\ft1{120}(k_2+11)^2+\ft{121}{120}$ under $T^{\rm eff}$, generalise to
a
continuous-momentum multi-scalar operator where $e^{i p\cdot X}$ has the
same dimension $\Delta=\ft12 p^\mu(p_\mu+2a_\mu)$.

\item{2)} If there is a two-scalar prime physical operator with momentum
$(k_1,\, 0)$.  This generalises to a discrete multi-scalar operator with
$p_\mu=0$ in the effective spacetime, and hence $\Delta=0$.

\item{3)} If there is a two-scalar prime physical operator with momentum
$(k_1,\, -22)$.  This generalises to a discrete multi-scalar operator with
$p_\mu=-2a_\mu$ in the effective spacetime, where $a_\mu$ is the constant
background-charge vector appearing in (5.14).  Again, this has conformal
dimension $\Delta=0$ as measured by $T^{\rm eff}$.
\medskip

     By taking the subset of two-scalar physical states that satisfies the
above conditions, and generalising them to the multi-scalar case, we obtain all
the multi-scalar states that are purely tachyonic in the effective spacetime.
The complete set of physical states, including excitations in the effective
spacetime, is then obtained by replacing the spacetime vertex operators $e^{i
p\cdot X}$ with arbitrary highest-weight operators with the same conformal
weight under $T^{\rm eff}$ given in (5.17).

     By looking in detail at all the prime physical operators (5.12) of the
two-scalar \w24 string, we find that the subset of physical operators that can
be generalised to the multi-scalar case is given by $X^m\, t_i$, $X^m\, u_i$,
$X^m\, v_i$, $X^m\,( Y^{-1}d_1)$, $X^m\,(Y d_3)$, $X^m\,( Y^{-1}d_2)$,
$X^m\,( Y d_4)$ (which
all generalise to continuous-momentum physical operators), and $X^m\, d_i$
(which generalise to discrete-momentum physical operators).  All the
fundamental
operators $t_i$, $u_i$ and $v_i$ given in (5.8)--(5.10) themselves occur in
pairs with conjugate $\varphi_2$ momenta (those with index $i=2r+2$ are
conjugate to those with index $i=2r+1$).  The operators $( Y^{-1}d_1)$ and
$( Y d_3)$ comprise a pair with conjugate $\varphi_2$ momenta: $(k_1,\,
k_2)=(-30,\, -30)$ and $(-30,\, 8)$.  Similarly, $( Y^{-1}d_2)$ and $( Y d_4)$
comprise a conjugate pair with momenta $(-24,\, -30)$ and $(-24,\, 8)$.   These
generalise to  continuous-momentum multi-scalar operators with $G=4$ at level
$\ell=3$, which we shall call $\tilde w_1$ and $\tilde w_2$ respectively.  Thus
we may re-express the cohomology of prime operators in the multi-scalar \w24
string using a fundamental basis comprising the multi-scalar generalisations of
(5.8)--(5.11), together with $\tilde w_i$.  The set of fundamental physical
operators for the multi-scalar \w24 string are therefore given by
$$
\eqalignno{
\tilde t_1&= c\, \del^2\gamma\, \del\gamma\, \gamma\,
e^{-\ft{30}{27}\alpha\varphi_1}\,e^{ip\cdot X};\qquad\Delta=1\ ,\cr
\tilde t_2&= c\, \del^2\gamma\, \del\gamma\, \gamma\,
e^{-\ft{24}{27}\alpha\varphi_1}\,e^{ip\cdot X};\qquad\Delta=1\ ,\cr
\tilde t_3&= c\, \del^2\gamma\, \del\gamma\, \gamma\,
e^{-\ft{28}{27}\alpha\varphi_1}\,e^{ip\cdot X};\qquad\Delta=\ft{14}{15}\ ,\cr
\tilde t_4&= c\, \del^2\gamma\,\del\gamma\, \gamma\,
e^{-\ft{26}{27}\alpha\varphi_1}\,e^{ip\cdot X};\qquad\Delta=\ft{14}{15}\ ,\cr
\cr
\tilde u_1&= c\,  \del\gamma\, \gamma\,
e^{-\ft{20}{27}\alpha\varphi_1}\,e^{ip\cdot X};\qquad\Delta=\ft{1}{3}\ ,\cr
\tilde u_2&= c\,  \del\gamma\, \gamma\,
e^{-\ft{18}{27}\alpha\varphi_1}\,e^{ip\cdot X};\qquad\Delta=\ft{3}{5}\ ,\cr
\tilde u_3&= c\,  \del\gamma\, \gamma\,
e^{-\ft{16}{27}\alpha\varphi_1}\,e^{ip\cdot X};\qquad\Delta=\ft{14}{15}
\ ,\cr
\tilde u_4&=c\, \Big( \del\varphi_1\, \del\gamma\, \gamma\, -\ft4{27}\alpha\,
\del^2\gamma\, \gamma\Big)\,
e^{-\ft{14}{27}\alpha\varphi_1}\,e^{ip\cdot X};\qquad\Delta=\ft{1}{3}\ ,\cr
\tilde u_5&=c\, \Big( \del\varphi_1\, \del\gamma\, \gamma\, -\ft2{15}\alpha\,
\del^2\gamma\, \gamma\Big)\,
e^{-\ft{18}{27}\alpha\varphi_1}\,e^{ip\cdot X};\qquad\Delta=-\ft{2}{5}\ ,\cr
\cr
\tilde v_1&=c\,\gamma\, e^{-\ft{10}{27}\alpha\varphi_1}\,e^{ip\cdot X};
\qquad\Delta=\ft{1}{3}\,\cr
\tilde v_2&=c\,\gamma\, e^{-\ft{8}{27}\alpha\varphi_1}\,e^{ip\cdot X};
\qquad\Delta=\ft{14}{15}\ ,\cr
\tilde v_3&=c\, \Big( \del\varphi_1\, \gamma\, -\ft8{27}\alpha\,
\del\gamma\Big)\,
e^{-\ft{6}{27}\alpha\varphi_1}\,e^{ip\cdot X};\qquad\Delta=\ft{3}{5}\ ,\cr
\tilde v_4&=c\, \Big(\beta\,\del\gamma\,\gamma
-\ft{4}{9}\alpha\,\del\varphi_1\,
\del\gamma +\ft1{18}\alpha\, \del^2\varphi_1\, \gamma +\ft34\,
(\del\varphi_1)^2 \,\gamma  +\ft3{10} \del^2\gamma\Big)\,
e^{-\ft{4}{27}\alpha\varphi_1}\,e^{ip\cdot X};\qquad\Delta=\ft{1}{3}\ ,\cr
\tilde v_5&=c\, \Big(\beta\,\del\gamma\,\gamma -\ft{35}{108}\alpha\,
\del\varphi_1\, \del\gamma +\ft5{36}\alpha\, \del^2\varphi_1\, \gamma +\ft58\,
(\del\varphi_1)^2 \,\gamma \Big)\,
e^{-\ft{6}{27}\alpha\varphi_1}\,e^{ip\cdot X};\qquad\Delta=-\ft{2}{5}\ ,\cr
\cr
\tilde w_1&=c\,\Big( 41\, (\del\varphi_1)^3\, \del^2\gamma\, \del
\gamma\,\gamma
+\ft{129}{10}\, \del\varphi_1\, \del^3\gamma\, \del^2\gamma\, \gamma
+\ft{89}{20}\,  \del\varphi_1\, \del^4\gamma\, \del\gamma\, \gamma \cr
&\qquad - \ft{12}5\,
\del^2\varphi_1\,  \del^3\gamma\, \del\gamma\, \gamma
-\ft{26}3\alpha\,(\del\varphi_1)^2 \, \del^3\gamma\, \del\gamma\, \gamma
+\ft{8}3\, \alpha\, \del^2\varphi_1\, \del\varphi_1 \, \del^2\gamma\,
\del\gamma\, \gamma\cr &\qquad -\ft{53}{45}\, \alpha\,
\del^3\gamma\, \del^2\gamma\, \del \gamma - \ft{11}{30}\, \alpha\,
\del^4\gamma\, \del^2\gamma\, \gamma -\ft{13}{450}\, \alpha\, \del^5\gamma\,
\del \gamma\, \gamma \Big)e^{-\ft{30}{27}\alpha \varphi_1}\, e^{i p\cdot
X};\qquad \Delta=-2\ ,\cr
\tilde w_2&=c\,\Big( 68\, (\del\varphi_1)^3\, \del^2\gamma\, \del
\gamma\,\gamma
+36\, \del\varphi_1\, \del^3\gamma\, \del^2\gamma\, \gamma -50\,
\del\varphi_1\, \del^4\gamma\, \del\gamma\, \gamma \cr
&\qquad-120\,
\del^2\varphi_1\,  \del^3\gamma\, \del\gamma\, \gamma + 15\, \del^3\varphi_1 \,
\del^2\gamma\, \del\gamma\, \gamma +\ft{340}3\, \alpha\, \del^2\varphi_1\,
\del\varphi_1 \, \del^2\gamma\, \del\gamma\, \gamma\cr
&\qquad -\ft{40}{3}\, \alpha\,
\del^3\gamma\, \del^2\gamma\, \del \gamma + \ft{10}{3}\, \alpha\,
\del^4\gamma\, \del^2\gamma\, \gamma +\ft{20}9\, \alpha\, \del^5\gamma\,
\del \gamma\, \gamma \Big)e^{-\ft{24}{27}\alpha \varphi_1}\, e^{i p\cdot
X};\qquad \Delta=-2\ ,\cr
\cr
\tilde d_1&=\Big( c\,\del\gamma\, \gamma -\ft{28}{15}\, \del^2\gamma\,
\del\gamma\, \gamma \Big)\, e^{-\ft{30}{27}\alpha\varphi_1};\qquad\Delta=0\
,\cr
\tilde d_2&=\Big( c\,\del\gamma\, \gamma -\ft{4}{15}\, \del^2\gamma\,
\del\gamma\, \gamma \Big)\, e^{-\ft{24}{27}\alpha\varphi_1};\qquad\Delta=0\
,\cr
\tilde d_3&=c\, \Big( \del^4\gamma\, \del^2\gamma\, \del\gamma\,\gamma
-\ft{10}9
\, \del\varphi_1\, \del^3\gamma\, \del^2\gamma\, \del\gamma\, \gamma \Big)
e^{-\ft{30}{27}\alpha\varphi_1}\, e^{-2i a\cdot X};\qquad\Delta=0\ ,\cr
\tilde d_4&=c\, \Big( \del^4\gamma\, \del^2\gamma\, \del\gamma\, \gamma
-\ft{8}9
\, \del\varphi_1\, \del^3\gamma\, \del^2\gamma\, \del\gamma\, \gamma \Big)
e^{-\ft{24}{27}\alpha\varphi_1}\, e^{-2i a\cdot X};\qquad\Delta=0\ .
&(5.18)\cr}
$$
Here $\Delta=\ft12p\cdot (p+2a)$ is the conformal weight of the spacetime
vertex operator $e^{ip\cdot X}$, as measured by $T^{\rm eff}$.  The complete
cohomology of prime operators in the multi-scalar \w24 string comprises
continuous-momentum operators $X^m\tilde t_i$ at ghost number $G=4-3m$,
$X^m\tilde u_i$ at  ghost number $G=3-3m$, $X^m\tilde v_i$ at ghost number
$G=2-3m$ and $X^m\tilde w_i$ at ghost number $G=4-3m$, together with
discrete-momentum operators $X^m \tilde d_1$ and $X^m \tilde d_2$ at $G=3-3m$
and $X^m \tilde d_3$ and $X^m \tilde d_4$ at $G=5-3m$.  The $\varphi_1$
momenta, spacetime weights $\Delta$, and level numbers can be obtained from
(5.18), and the mass-shell formula (5.2).

     We see from (5.18) that the continuous-momentum fundamental operators
$\tilde t_i$, $\tilde u_i$, $\tilde v_i$ and $\tilde w$ have conformal weights
$\Delta=\{1,\, \ft{14}{15},\, \ft35,\, \ft13,\, -\ft25,\, -2\}$, with
multiplicities $\{2,\, 4,\, 2,\, 4,\, 2,\, 2\}$ respectively. The reason for
the
doubling of the multiplicities of the $\Delta=\ft{14}{15}$ and $\Delta=\ft13$
operators can be understood as follows:    All the continuous-momentum physical
operators have the form $c\, H(\varphi_1,\beta,\gamma)\, e^{ip\cdot X}$, and
thus the operators $H(\varphi_1,\beta,\gamma)$ have weights $h$ that are
conjugate to $\Delta$, in the sense that $h=1-\Delta$.  The weights $h$ are in
fact those of the primary fields of the $c=\ft45$ unitary Virasoro minimal
model, namely $h=\{0, \ft1{15},\, \ft25,\, \ft23,\, \ft75,\,  3\}$.  It was
shown in [15] that indeed the $H(\varphi_1,\beta,\gamma)$ are highest-weight
operators of the $c=\ft45$ Virasoro minimal model.  As was shown in [16], the
operators $H(\varphi_1,\beta,\gamma)$ also correspond to states in the lowest
unitary $W_3$ minimal model, which also has $c=\ft45$.  This $W_3$ algebra is
generated by the spin-2 current $T_{\varphi_1}+T_{\beta,\gamma}$ and the spin-3
primary current $U$ given in (5.6).  The $H(\varphi_1,\beta,\gamma)$ operators
with Virasoro weights $h=\{0,\, \ft1{15},\, \ft25,\, \ft23\}$ are
highest-weight
under the spin-3 generator $U$, whilst those with weights $h=\ft75$ and $h=3$
are $W_3$ descendants [16].  Of the $W_3$ highest-weight operators, those with
$h=0$ and $\ft25$ have zero weight under the spin-3 current $U$, whilst those
with $h=\ft1{15}$ and $\ft23$ have non-zero weights under $U$.  There are in
fact two different highest-weight states in each of these two cases, one with
positive, and the other with equal and opposite negative, weight under $U$.
Thus the multiplicity of fundamental operators associated with these two
conformal weights should be double the multiplicity for all the other conformal
weights.

     The $\Delta=-2$ physical operators $\tilde w_1$ and $\tilde w_2$ in (5.18)
are examples that correspond to $W$-descendant operators in the $W_3$
minimal model.  For example we find that $\tilde w_1$ can be written as
$$
\tilde w_1\propto \Big( U_{-3} +\ft1{5824\alpha}(1775\, U_{-1} U_{-2} -605\,
U_{-2}U_{-1} ) +\ft{75}{18928\alpha^2}\, (U_{-1})^3\Big) \, \tilde t_1\ ,
\eqno(5.19)
$$
where $U$ is the spin-3 primary current of the $W_3$ minimal model, given in
(5.6), and $\tilde t_1$ is a $\Delta=1$ tachyon, given in (5.18).  In fact
there is a freedom to add a BRST-trivial term to $\tilde w_1$ in (5.18), whilst
preserving its factorised form, and only for one particular choice of the
associated free parameter, namely the choice we have adopted, can $\tilde w_1$
be written as a descendant operator.  Similarly, we may write the operator
$\tilde w_2$ in (5.18) as a $W_3$ descendant.  In this case, the BRST-trivial
term that must be included in $\tilde w_2$ in order to allow it to be expressed
as a $W_3$ descendant results in rather complicated coefficients, and so we
have left $\tilde w_2$ in a simpler form.  Thus up to BRST-trivial terms,
we find that may write $\tilde w_2$ as
$$
\Big( U_{-3} -\ft1{128\alpha} (211\, U_{-1} U_{-2} +1076\, U_{-2}U_{-1} )
-\ft{519}{64\alpha^2}\, (U_{-1})^3\Big) \, \tilde t_2\ .\eqno(5.20)
$$

\np
\noindent{\bf 6. Critical $W_{2,s}$ Strings}
\bigskip

     It is of interest to consider how the results of the previous sections
might be extended to the case of $W_{2,s}$ strings for general values of $s$.
A ``critical'' BRST operator of the form (2.1)--(2.7) has been explicitly
constructed for $s=3$ (the $W_3$ case [6]), and $s=4$, 5 and 6 [15].  It is
expected that such BRST operators will exist for arbitrary values of $s$.

     Let us consider first the possible generalisation to arbitrary $s$ of the
construction of the non-critical $W_{2,4}$ string in section 3. In this
construction, the critical $W_{2,s}$ BRST operator is extended by introducing
additional matter currents that satisfy the $W_{2,s}$ algebra.  As we discussed
in section 3, such an algebra, the $W\!B_2$ algebra, exists at arbitrary
central charge in the $s=4$ case.  A similar situation obtains for $s=6$, in
which case the algebra in question is $W\!G_2$ [17,18].  Thus we expect that a
construction of a non-critical $W_{2,6}$ BRST operator along the lines of
section 3 should be possible in principle.  In practice, the complexity of the
system would make an explicit construction quite difficult.  For the case of
$s=5$, a $W_{2,5}$ algebra exists only at certain discrete values of the
central charge.  In principle, one could imagine that a non-critical $W_{2,5}$
BRST operator could be built, in which the matter system realises the $W_{2,5}$
algebra at one or another of these discrete values of central charge.  Again,
the system is a complicated one, and it would be quite difficult to carry out
the construction explicitly.  For $s\ge 7$, it is again known that $W_{2,s}$
algebras could exist at most for discrete values of central charge.  The
complexities of the systems would be even greater in these cases.

     Turning now to the critical $W_{2,s}$ BRST operators, it is of interest
to consider generalising the $W_{2,4}$ cohomology discussion of section 5.  As
we shall see, it turns out that there are new features that arise when $s\ge 5$
that complicate the discussion considerably.  It was already conjectured in
[15] that the physical states of the multi-scalar $W_{2,s}$ string all have
the form of effective Virasoro physical states tensored with primary or
descendant operators of the lowest unitary $W_{s-1}$ minimal model, which has
central charge $c={2(s-1) \over (s+2)}$.  This was investigated further in
[16],
where strong supporting evidence for the conjecture was obtained.  Since the
minimal model has $c\ge1$ when $s\ge5$, it follows that if one decomposes the
operators under the Virasoro subalgebra, an infinite number of primary Virasoro
operators will arise.  This has the consequence for the multi-scalar $W_{2,s}$
string theory that there will be an infinite number of effective-spacetime
physical sectors, with weights $\Delta$ of the form $\Delta_i=1-h_W-N_i$, where
$h_W$ takes values in the (finite) list of conformal weights of the primary
operators of the lowest $W_{s-1}$ minimal model, and the integers $N_i$ take
infinite sets of values, corresponding to the weights $h_W+N_i$ of operators
that are $W_{s-1}$ descendants but that are still Virasoro primaries [16].
Thus
if one looks for the analogue of the set of ``basic'' physical operators (5.18)
for the higher $W_{2,s}$ strings with $s\ge5$, the set will itself be
infinite.  (This did not happen for $s=3$ or $s=4$, since the relevant minimal
models had central charges $c=\ft12$ and $c=\ft45$ respectively, and thus the
lists of Virasoro primaries were finite.)

     A related complication when $s\ge 5$ is that it is no longer the case that
the $\varphi_2$ momenta of all physical states in the two-scalar $W_{2,s}$
string are rational multiples of the background charge  $a$.  Let us focus
in particular on the subset of two-scalar states that can be generalised to
continuous-momentum states in the multi-scalar string.  We have looked
at the $W_{2,5}$ string, for which the background charges are given by
$\alpha^2=\ft{121}6$ and $a^2=2$ [15], in some detail, and we find that
although the $\varphi_1$ momentum seems to be quantised in units of
$\ft1{22}\,\alpha$, the $\varphi_2$ momenta of physical states can have one of
two possible forms; $p_2=\ft14\, a\, k_2$ or $p_2=-a +\ft{a}{2\sqrt{3} }\,
k_2$,
where $k_2$ is an integer.  The reason for this can be seen by looking at the
relation  between the conformal weights $\Delta=-\ft12 p_2(p_2+2a)$ of physical
states of the $W_{2,s}$ string under $T_{\varphi_2}$, and the conformal weights
$h=1-\Delta$ of the primary operators of the associated $W_{s-1}$ minimal
model.  Writing
$$
p_2=q_2\, a\eqno(6.1)
$$
in each case, we find that the relations are as follows:
$$
\eqalignno{
W_{2,3}&: \qquad\qquad (q_2+1)^2=\ft1{49}(48\, h+1)\ ,&(6.2a)\cr
W_{2,4}&: \qquad\qquad (q_2+1)^2=\ft1{121}(120\, h+1)\ ,&(6.2b)\cr
W_{2,5}&: \qquad\qquad (q_2+1)^2=h \ ,&(6.2c)\cr
W_{2,6}&: \qquad\qquad (q_2+1)^2=\ft1{167}(168\, h-1)\ ,&(6.2d)\cr}
$$
The weights $h$ for each case are as follows [15,16]:
$$
\eqalignno{
W_{2,3}&:\qquad h=\{0,\, \ft1{16},\, \ft12 \}\ ,&(6.3a)\cr
W_{2,4}&:\qquad h=\{0,\, \ft1{15},\, \ft25,\, \ft23;\, \ft75,\, 3 \}\ ,
&(6.3b)\cr
W_{2,5}&:\qquad h=\{0,\, \ft1{16},\, \ft1{12},\, \ft13,\, \ft9{16},\,
\ft34,\, 1;\, \ft43,\, \ft{25}{16},\, \ft{25}{12},\, \ft{49}{16},\, \cdots
\}\,&(6.3c)\cr
W_{2,6}&:\qquad h=\{0,\, \ft2{35},\, \ft3{35},\, \ft27,\, \ft{17}{35},\,
\ft{23}{35},\, \ft45,\, \ft67,\, \ft65;\, \ft97,\, \ft{52}{35},\, \ft{58}{35},
\, \ft{13}7,\, \ft{73}{35},\, \ft{87}{35},
\, \ft{20}{7},\, \ft{16}{5},\,\cdots\}\,&(6.3d)\cr}
$$
In each case, the weights to the left of the semicolon are those of primary
operators of the $W_{s-1}$ minimal model, whilst those  after the
semicolon (if any) are the weights of $W_{s-1}$ descendants that are Virasoro
primaries. For the $W_{2,3}=W_3$ string, we see that for each value of $h$ in
(6.3$a$), the right-hand side of (6.2$a$) has a rational square root.  The same
is true for (6.2$b$) when one substitutes the $h$ values (6.3$b$) of the
$W_{2,4}$ string.   However,  for the $W_{2,5}$ string we see that only some of
the $h$ values (6.3$c$) give rational roots for $q_2$,
$$
q_2=\ft14\, k_2\ ,\eqno(6.4)
$$
whilst the remaining $h$ values give $q_2$ of the form
$$
q_2=-1 + \ft1{2\sqrt{3}}\, k_2\ ,\eqno(6.5)
$$
where $k_2$ is an integer in each case.  Finally, for the $W_{2,6}$ string
we see from (6.3$d$) there are many irrationally-related values for $q_2$ (the
occurrence of the prime number 167 in the denominator of (6.2$d$), unlike the
perfect square integers of the previous cases, makes irrationality much more
probable).

      One consequence of the occurrence of physical states with
irrationally-related $\varphi_2$ momenta is that there cannot exist universal
$Y$-type operators that can normal order with any physical state to give
other physical states. On the other hand, $X$-type operators (which have zero
momentum in the $\varphi_2$ direction), can still be expected to exist, and
they can still normal order with all physical operators.

     Let us consider first the $W_{2,5}$ string. By analogy to these previous
cases, we expect that the $X$ operator should have as its associated screening
current the simple expression
$$
S_X=\del^4\beta\, \del^3\beta\,\del^2\beta\, \del\beta\, \beta\,
e^{\ft{12}{11}\alpha\varphi_1}\ .\eqno(6.6)
$$
Thus the $X$ operator itself will have ghost number $G=-4$.  Checking that
$[Q_B, S_X]=\del X$ where $Q_B$ is the BRST operator given in [15] is a
somewhat
non-trivial calculation, which we have not been able to
complete.\footnote{$^*$}{\tenfoot We do, however, know that there exists a
$G=0$
physical operator $x$ at level $\ell=15$, whose screening current is given by
$S_x=\beta\, e^{\ft2{11}\alpha\varphi_1}$ [15].  By acting with five of these
screening currents (and a ghost booster) on the physical operator $x$, one
should, in principle, be able to obtain the required $X$ operator.  Thus the
known existence of $x$ provides supportive evidence for the existence of
$X$.}  However, the inverse of $X$, at level $\ell=1$ with $G=4$, is easily
computed:
$$
X^{-1}=\Big( c\, \del^2\gamma\,\del\gamma\,\gamma -\ft5{528}\alpha\,
\del^3\gamma\, \del^2\gamma\, \del\gamma\, \gamma\Big)\,
e^{-\ft{12}{11}\alpha\varphi_1}\ . \eqno(6.7)
$$
As far as a $Y$-type operator is concerned, one might hope that since, for the
$W_{2,5}$ string, there appear to be just the two possible sequences of
$q_2$ momenta given by (6.4) and (6.5), there might at least exist an
invertible $Y$ operator that could normal order with physical operators whose
$q_2$ momenta are given by (6.4).  For physical operators of this kind, which
we
shall call ``rational'' operators, the level number is related to the momentum
$(p_1,\, p_2)=(\ft1{22}\alpha\, k_1,\, \ft14 a\, k_2)$ by the mass-shell
condition
$$
(k_1+22)^2 + 3(k_2 + 4)^2 =4(12\ell+1)\ .\eqno(6.8)
$$
Necessary conditions that should be satisfied by a candidate $Y$ operator are
the following.  Firstly its momentum should satisfy (6.8) for integer $k_1$,
$k_2$ and $\ell$.  Secondly $(-k_1,\, -k_2)$, which would be the momentum of
the $Y^{-1}$ operator, should also satisfy (6.8) for some other integer
$\ell'$.
Thirdly, the $Y$ operator should give integer-degree poles in its operator
product with any other rational physical operator. These conditions imply that
$(k_1,\, k_2)$ should have the form $(k_1,\, k_2)=(12m,\, 4n)$, where the
integers $m$ and $n$ are either both odd, or both even.  Of course these are
only necessary conditions for the existence of an invertible $Y$ operator, and
unfortunately the physical-state conditions are too complicated for us to be
able to determine whether any such operator actually exists.  It is also
conceivable that the fact that there are also ``irrational'' physical operators
whose $\varphi_2$ momenta satisfy (6.5), which certainly could not normal order
with any $Y$, is an indication that no invertible $Y$ operator can exist.

     It was shown in [11] for the $W_3$ string, and in section 5 of this paper
for the \w24 string, that in the multi-scalar case the entire cohomology can be
constructed by acting with arbitrary powers of the relevant $X$ operator on a
set of basic multi-scalar states.  We may expect therefore that the same should
be true for the $W_{2,5}$ string, and thus the lack of a $Y$ operator need not
be an obstacle to constructing the cohomology in this multi-scalar case.  Of
course, as we discussed earlier, there are an infinite number of different
matter sectors in the $W_{2,5}$ string, corresponding to an infinite number of
different effective-spacetime weights $\Delta$.  Consequently, the set of basic
operators $u_i$, from which all prime physical operators will be obtained as
$X^m\, u_i$, will itself be infinite.  However, as was shown in [16], this
infinite set can be understood as arising from decomposing the primary
operators
of the $c=1$ $W_4$ algebra into $W$ descendants that are still Virasoro
primaries, and so by this means the infinite set $\{u_i\}$ can be generated in
principle from just a finite set of physical operators with effective-spacetime
weights $\Delta=1-h_W$, where $h_W=\{0,\, \ft1{16},\, \ft1{12},\, \ft13,\,
\ft9{16},\, \ft34,\, 1\}$ is the set of conformal weights of the primary fields
of the $c=1$ $W_4$ minimal model.  Furthermore, we expect that the
multiplicities
for the basic set of operators for the $h_W$ values listed above should be
$\{2,\, 4,\, 2,\, 2,\, 4,\, 4,\, 2\}$.  The reason for the doubling of those
with
conformal weights $h_W=\{ \ft1{16},\, \ft9{16},\, \ft34\}$ is the same as we
discussed in section 5 for certain of the \w24 basic states, namely that the
corresponding primary operators of the $W_4$ minimal model have non-zero
weights
under the spin-3 current, and thus occur in $\pm$ pairs.  The 20 basic
continuous-momentum physical operators, together with their effective-spacetime
weights $\Delta=1-h_W$, are:
$$
\eqalignno{
t_1&=c\, \del^3\gamma\, \del^2\gamma\, \del\gamma\, \gamma\, e^{-\ft{24}{22}
\alpha \varphi_1}\, e^{i p\cdot X}\, ;\qquad
t_2=c\, \del^3\gamma\, \del^2\gamma\, \del\gamma\, \gamma\, e^{-\ft{20}{22}
\alpha \varphi_1}\, e^{i p\cdot X},\quad \Delta=1 \cr
t_3&=c\, \del^3\gamma\, \del^2\gamma\, \del\gamma\, \gamma\, e^{-\ft{23}{22}
\alpha \varphi_1}\, e^{i p\cdot X}\, ;\qquad
t_4=c\, \del^3\gamma\, \del^2\gamma\, \del\gamma\, \gamma\, e^{-\ft{21}{22}
\alpha \varphi_1}\, e^{i p\cdot X},\quad \Delta=\ft{15}{16} \cr
u_1&=c\,  \del^2\gamma\, \del\gamma\, \gamma\, e^{-\ft{15}{22}
\alpha \varphi_1}\, e^{i p\cdot X}\, ;\qquad
w_1=c\,  \gamma\, e^{-\ft{5}{22}
\alpha \varphi_1}\, e^{i p\cdot X},\quad \Delta=\ft{15}{16} \cr
t_5&=c\, \del^3\gamma\, \del^2\gamma\, \del\gamma\, \gamma\, e^{-
\alpha \varphi_1}\, e^{i p\cdot X}\, ;\qquad
v_1=c\, \del\gamma\, \gamma\, e^{-\ft{10}{22}
\alpha \varphi_1}\, e^{i p\cdot X},\quad \Delta=\ft{11}{12} \cr
u_2&=c\,\del^2\gamma\, \del\gamma\, \gamma\, e^{-\ft{16}{22}
\alpha \varphi_1}\, e^{i p\cdot X}\, ;\qquad
w_2=c\,\big( \del\gamma -\ft{12}{55}\alpha\,\del\varphi_1\, \gamma\big)\,
e^{-\ft{4}{22} \alpha \varphi_1}\, e^{i p\cdot X},\quad \Delta=\ft{2}{3}\cr
u_3&=c\, \del^2\gamma\, \del\gamma\, \gamma\, e^{-\ft{17}{22}
\alpha \varphi_1}\, e^{i p\cdot X}\, ;\qquad
v_2=c\, \del\gamma\, \gamma\, e^{-\ft{11}{22}
\alpha \varphi_1}\, e^{i p\cdot X},\quad \Delta=\ft7{16} \cr
v_3&=c\,\big( \del^2\gamma\,\gamma -\ft{24}{55}\alpha\, \del\varphi_1\,
\del\gamma\, \gamma \big)\, e^{-\ft{9}{22}\alpha \varphi_1}\, e^{i p\cdot X},
\quad \Delta=\ft7{16}\cr
w_3&=c\, \big( \del^2\gamma -\ft{12}{11}\alpha\, \del\varphi_1\, \del\gamma
+\ft{12}5\, (\del\varphi_1)^2\, \gamma +\ft6{55}\alpha\, \del^2\varphi_1\,
\gamma +3\, \beta\, \del\gamma\, \gamma \big)\, e^{-\ft{3}{22}\alpha \varphi_1}
\, e^{i p\cdot X},\quad \Delta=\ft7{16}\cr
u_4&=c\, \del^2\gamma\, \del\gamma\,
\gamma\, e^{-\ft{18}{22} \alpha \varphi_1}\, e^{i p\cdot X}\, ;\qquad  w_4=c\,
\gamma\, e^{-\ft{6}{22} \alpha \varphi_1}\, e^{i p\cdot X},\quad \Delta=\ft14
\cr
u_5&=c\,\big(  \del^3\gamma\,\del\gamma\, \gamma -\ft{36}{55}\alpha\,
\del\varphi_1\,
 \del^2\gamma\, \del\gamma\, \gamma\big)\, e^{-\ft{14}{22}\alpha \varphi_1}
\, e^{i p\cdot X},\quad \Delta=\ft14\cr
w_5&=c\,\Big( \del^3\gamma +36\, (\del\varphi_1)^2\, \del\gamma -\ft{36}5\,
\del^2\varphi_1\, \del\varphi_1\, \gamma -\ft{36}{11}\alpha\, \del\varphi_1\,
\del^2\gamma\cr
&\qquad-\ft{144}{55}\alpha\, (\del\varphi_1)^3\, \gamma
+\ft{18}{11}\alpha\, \del^2\varphi_1\, \del\gamma -\ft6{55}\alpha\,
\del^3\varphi_1\, \gamma \cr
&\qquad+12\, \beta\, \del^2\gamma\, \gamma
 -21\, \del\beta\,
\del\gamma\, \gamma -\ft{108}{11}\alpha\, \del\varphi_1\, \beta\, \del\gamma\,
\gamma \Big)\,e^{-\ft{2}{22}\alpha \varphi_1}
\, e^{i p\cdot X},\quad \Delta=\ft14\cr
v_4&=c\,  \del\gamma\, \gamma\, e^{-\ft{12}{22}
\alpha \varphi_1}\, e^{i p\cdot X}, \quad \Delta=0  &(6.9) \cr
v_5&=c\, \Big(\del^3\gamma\, \gamma +\ft92\, \del^2\gamma\, \del\gamma
-\ft{18}{11}\alpha\, \del\varphi_1\, \del^2\gamma\, \gamma +\ft{36}5\,
(\del\varphi_1)^2\, \del\gamma\, \gamma +\ft{18}{55}\alpha\, \del\gamma\,
\gamma \Big)\, e^{-\ft{8}{22}\alpha  \varphi_1} \, e^{i p\cdot X},\quad
\Delta=0\cr}
$$
The cohomology of continuous-momentum prime physical
operators for the $W_{2,5}$ string, with $\Delta=\{1,\, \ft{15}{16},\,
\ft{11}{12},\, \ft23,\, \ft7{16},\, \ft14,\, 0\}$, is then given by $X^m t_i$
at
$G=5-4m$, $X^m u_i$ at $G=4-4m$,  $X^m v_i$ at $G=3-4m$, and $X^m w_i$ at
$G=2-4m$.  The rest of the continuous-momentum prime physical operators, with
$\Delta\le 0$, are then obtainable as $W_4$ descendants [16].  Some of the
higher-level physical operators found in [15] provide non-trivial consistency
checks of the existence of the invertible $X$ operator.

     There will in addition be physical operators with discrete momentum in
the effective spacetime.  These will come from four basic discrete operators
at $\ell=1$:
$$
\eqalign{
d_1&=\Big( c\, \del^2\gamma\,\del\gamma\,\gamma -\ft5{528}\alpha\,
\del^3\gamma\, \del^2\gamma\, \del\gamma\, \gamma\Big)\,
e^{-\ft{12}{11}\alpha\varphi_1}, \quad \Delta=0\cr
d_2&=\Big( c\, \del^2\gamma\,\del\gamma\,\gamma -\ft{35}{176}\alpha\,
\del^3\gamma\, \del^2\gamma\, \del\gamma\, \gamma\Big)\,
e^{-\ft{10}{11}\alpha\varphi_1}, \quad \Delta=0\cr
d_3&=c\, \Big(\del^5\gamma\, \del^3\gamma\, \del^2\gamma\, \del\gamma\, \gamma
-\ft{12}{11}\alpha\,  \del\varphi_1\,\del^4\gamma\,
\del^3\gamma\,\del^2\gamma\,
\del\gamma\, \gamma \Big) \,e^{-\ft{12}{11}\alpha\varphi_1}\, e^{-2ia\cdot X},
\quad \Delta=0\cr
d_4&=c\, \Big(\del^5\gamma\, \del^3\gamma\, \del^2\gamma\, \del\gamma\, \gamma
-\ft{10}{11}\alpha\,  \del\varphi_1\,\del^4\gamma\,\del^3\gamma\,
\del^2\gamma\,
\del\gamma\, \gamma \Big) \,e^{-\ft{10}{11}\alpha\varphi_1}\, e^{-2ia\cdot X},
\quad \Delta=0\cr}\eqno(6.10)
$$
Thus there will be discrete prime operators $X^m d_1$ and $X^m d_2$ at
$G=4-4m$, and $X^m d_3$ and $X^m d_4$ at $G=6-4m$.

     It is worth noting that the set of 20 basic continuous-momentum operators
given in (6.9), comprises five each at ghost numbers $G=5$, 4, 3 and 2, namely
$\{t_i\}$, $\{u_i\}$, $\{v_i\}$ and $\{w_i\}$.  Similarly, if we consider the
basic continuous-momentum operators for the \w24 string, given in (5.18), and
restrict attention to the ones associated with the primary fields of the
$c=\ft45$ $W_3$ minimal model ({\it i.e.}\ exclude the $\Delta=-\ft25$ and
$\Delta=-2$ operators in (5.18), as well as the $\Delta=0$ discrete
operators), then there are a total of twelve, comprising four at each of the
ghost numbers $G=4$, 3 and 2.  In the $W_3$ string case, discussed in [11],
there are a total of six basic continuous-momentum operators, comprising three
at each of the ghost numbers $G=3$ and 2.  For the $W_{2,s}$ string, we can
expect that there should be $s(s-1)$ basic continuous-momentum operators
associated with the $W_{s-1}$ primary fields, with $s$ of them at each of the
ghost numbers $G=s$, $s-1$, $\ldots,\ 2$.

     The above considerations presumably give the complete cohomology of
prime physical operators in the multi-scalar critical $W_{2,5}$ string.
Although we have not been able to construct the enlarged cohomology of the
two-scalar case, it is certain that there are then additional physical
operators that arise, which cannot generalise to the multi-scalar case.  One
example that we have found, at level $\ell=6$ and ghost number $G=2$, is
$$
\eqalign{
V&=\Big( c\, \gamma +\ft{35}8\, \del\varphi_1\, \del^2\gamma\, \gamma
-\ft{1085}{48}\, \del^2\varphi_1\, \del\gamma\, \gamma -\ft{35}{22}\alpha\,
(\del\varphi_1)^2\, \del\gamma\, \gamma
-\ft{35}{88}\alpha\, (\del\varphi_2)^2\, \del\gamma\, \gamma\cr
&\quad -\ft{105}{88}
\alpha\, \del b\, c\, \del\gamma\, \gamma -\ft{455}{176}\alpha \,
\del^2\gamma\, \del\gamma +\ft{35}{48}\alpha\, \del^3\gamma\, \gamma
 -\ft{35}{16}a\, \del\varphi_1\, \del\varphi_2\, \del\gamma\, \gamma
-\ft{35}{44}a\, \alpha\, \del\varphi_2 \, b\, c\, \del\gamma\, \gamma \cr
&\quad +
\ft{35}{176}a\, \alpha\, \del\varphi_2\, \del^2\gamma\, \gamma -\ft{35}{176}a\,
\alpha\, \del^2\varphi_2\, \del\gamma\, \gamma \Big)\,
e^{-\ft{6}{11}\alpha\varphi_1+a\varphi_2}\ . \cr}\eqno(6.11)
$$

     The cohomology of the multi-scalar $W_{2,6}$ string should also be
obtainable by using these methods.  We expect that there should be an $X$
operator at $\ell=66$, corresponding to the screening current
$$
S_X=\del^5\beta\, \del^4\beta\, \del^3\beta\,\del^2\beta\, \del\beta\, \beta\,
e^{\ft{14}{13}\alpha\varphi_1}\ ,\eqno(6.12)
$$
where $\alpha^2=\ft{845}{28}$ in this case.  Checking that $[Q_B,S_X]=\del X$,
with $Q_B$ given in [15] for the $W_{2,6}$ string, would be extremely
complicated.\footnote{$^*$}{\tenfoot As in the case of the $W_{2,5}$ string
discussed earlier, we may use the known existence of a $G=0$ physical operator
$x$ at level $\ell=21$ in the $W_{2,6}$ string,  with associated screening
current $S_x=\beta\, e^{\ft2{13}\alpha\varphi_1}$ [15], to argue that the $X$
operator should arise at $\ell=66$ by acting with six screening currents $S_x$
and a ghost booster on the operator $x$.}  The $X$ operator has ghost number
$G=-5$.  The inverse $X^{-1}$ is, as usual, a simple level $\ell=1$ physical
operator, with $G=5$ in this case, given by
$$
X^{-1}= \Big( c\, \del^3\gamma\, \del^2\gamma\, \del\gamma\, \gamma
-\ft{396}{1225}\alpha\,  \del^4\gamma\, \del^3\gamma\, \del^2\gamma\,
\del\gamma\, \gamma\Big)\, e^{-\ft{14}{13}\alpha \varphi_1} \ .\eqno(6.13)
$$
In accordance with the general observations made above, we expect that the
continuous-momentum physical operators should be given by $X^m$
acting on a total of 30 basic operators, comprising six at each of the ghost
numbers $G=6$, 5, 4, 3 and 2.  This will give the full cohomology of operators
with $\Delta$ weights corresponding to the $W_5$ primary conformal weights
$h=1-\Delta$ in (6.3$d$).  The rest of the continuous-momentum physical
operators, with spacetime weights $\Delta=1-h$ corresponding to $h$ values to
the right of the semicolon in (6.3$d$), will arise as $W_5$ descendants [16].

\np
\noindent{\bf 7. Conclusion}
\bigskip

     In this paper, we have studied in detail aspects of some higher-spin
string
theories which have local spin-2 and spin-$s$ symmetries on the
two-dimensional worldsheet.   Such theories, which we call $W_{2,s}$ strings,
apparently exist for all $s\ge3$, although owing to the increasing
complexity at higher values of $s$, they have been explicitly constructed only
for $s=3$, 4, 5 and 6 [15].  The multi-scalar $W_{2,s}$ string has a physical
spectrum that is related to a tensor product of effective Virasoro string
states times the primary fields of the lowest unitary $W_{s-1}$ minimal model
[15,16].

     For most of the paper, we have concentrated on the case of the $W_{2,4}$
string.  This is the simplest generalisation of the $W_{2,3}$ case, which is
just the $W_3$ string.  After a brief review of the construction of the
critical $W_{2,s}$ string in section 2, we then presented new results in
section 3 in which we obtained the ``non-critical'' $W_{2,4}$ string.  This is
a generalisation of the non-critical $W_3$ string constructed in [12,13], in
which a nilpotent BRST operator is obtained that describes the coupling of
$W\!B_2$ matter to the pure $W\!B_2$ gravity of the critical $W_{2,4}$
string.    In section 4, we studied the spectrum of the theory and obtained
some
of the physical states, including ground-ring generators with ghost number
$G=0$
at levels $\ell=10$ and 11.  Obtaining a complete solution to the cohomology of
physical states would be very complicated in the non-critical \w24 string, but
in section 5 we were able to solve the problem for the simpler case of the
critical \w24 string, using the methods developed in [11].  Similar
considerations for the higher $W_{2,s}$ strings reveal the emergence of new
features, seemingly related to the fact that for $s\ge5$ the multi-scalar
states
are associated with $W_{s-1}$ minimal models whose primary fields decompose
into
infinite numbers of Virasoro primaries associated with non-unitary Virasoro
models.

    As well as the higher-spin BRST operators that we have been considering in
this paper, there are others that were found in [15] in the case of $s=4$ and
$s=6$.  These other BRST operators appear to be associated with non-unitary
string theories, in the sense that in the multi-scalar case the conformal
weights of the effective-spacetime physical operators can exceed 1, and thus
the longitudinal modes of excited states can have negative norms.  However, in
the two-scalar case, there are no longitudinal modes, and the norms of all
physical states can be expected to be positive.  It may be that there are some
interesting new features in these theories.

     Finally, we remark that in an interesting recent paper, it was proposed
that one can consider hierarchies of string theories in which, for example, the
$N=0$ bosonic string is viewed as a special vacuum of the $N=1$ superstring,
which in turn is viewed as a special vacuum of the $N=2$ string [25].  Another
possible route is to view the bosonic string as being embedded in a hierarchy
of
$W$--string theories [25,26].  Indeed, the realisation of the critical $W_3$
algebra, with $c=100$, in terms of an energy-momentum tensor $T^{\rm eff}$ with
$c=\ft{51}2$ and a scalar field $\varphi_1$, is very reminiscent of   the
procedure in [25] in which the critical $N=1$ super-Virasoro algebra, with
$c=15$, is realised in terms of an energy-momentum tensor $T_m$ with $c=26$ and
a $(b_1,\, c_1)$ anticommuting ghost system with spins $(\ft32,\, -\ft12)$.  In
both cases, the physical states of the corresponding $W_3$ string or $N=1$
superstring turn out to be rather trivial, in the sense that they are really
just states of a bosonic string.  However, we know that there are other
realisations of the $N=1$ super-Virasoro algebra that give a completely
different physical spectrum, namely that of the usual $N=1$ string.  It may be
that there correspondingly exist other realisations of $W$ algebras that  can
give rise to a more interesting spectrum of physical states. In view of these
points, it is of interest to explore the possible higher-spin string theories,
and their hierarchical structures, in more detail.

\bigskip\bigskip
\singlespace
\centerline{\bf REFERENCES}
\frenchspacing
\bigskip

\item{[1]}V.A. Fateev and A.B. Zamolodchikov, {\sl Nucl. Phys.} {\bf B280}
(1987) 644.

\item{[2]} C.N. Pope, L.J. Romans and K.S. Stelle, {\sl Phys. lett.} {\bf
B268} (1991) 167; {\sl Phys. lett.} {\bf B269} (1991) 287;

\item{[3]} A. Bilal and J.-L. Gervais, {\sl Nucl Phys.} {\bf B314} (1989)
646.

\item{[4]}S.R. Das, A. Dhar and S.K. Rama, {\sl Mod. Phys. Lett.}
{\bf A6} (1991) 3055; {\sl Int. J. Mod. Phys.} {\bf A7} (1992) 2295.

\item{[5]}C.N. Pope, L.J. Romans, E. Sezgin and K.S. Stelle,
{\sl Phys. Lett.} {\bf B274} (1992) 298.

\item{[6]}J. Thierry-Mieg, {\sl Phys. Lett.} {\bf B197} (1987) 368.

\item{[7]}F. Bais, P. Bouwknegt, M. Surridge and K. Schoutens, {\sl Nucl.
Phys} {\bf B304} (1988) 348.

\item{[8]}H. Lu, C.N. Pope, S. Schrans and K.W.
Xu,  {\sl Nucl. Phys.} {\bf B385} (1992) 99.

\item{[9]}H. Lu, C.N. Pope, S. Schrans and X.J. Wang, {\sl Nucl. Phys.} {\bf
B403} (1993) 351.

\item{[10]}H. Lu, C.N. Pope, S. Schrans and X.J. Wang,  ``On the spectrum
and scattering of $W_3$ strings,'' preprint CTP TAMU-4/93, KUL-TF-93/2,
hep-th/9301099, to appear in {\sl Nucl. Phys.} {\bf B.}

\item{[11]}H. Lu, C.N. Pope, X.J. Wang and K.W. Xu,  ``The complete
cohomology of the $W_3$ string,'' preprint CTP TAMU-50/93, hep-th/9309041.

\item{[12]}M. Bershadsky, W. Lerche, D. Nemeschansky and N.P. Warner,
{\sl Phys. Lett.} {\bf B292} (1992) 35;
``Extended $N=2$ superconformal structure of gravity and $W$-gravity
coupled to matter,'' USC-92/021, CERN-TH.6694/92.

\item{[13]}E. Bergshoeff, A. Sevrin and X. Shen, {\sl Phys. Lett.} {\bf
B296} (1992) 95.

\item{[14]}A. Anderson, B.E.W. Nilsson, C.N. Pope and K.S. Stelle, to appear.

\item{[15]}H. Lu, C.N. Pope and X.J. Wang,  ``On higher-spin
generalisations of string theory,'' preprint CTP TAMU-22/93,
hep-th/9304115, to appear in {\sl Int. J. Mod. Phys.} {\bf A}.

\item{[16]}H. Lu, C.N. Pope, K. Thielemans and X.J. Wang, ``Higher-spin
strings and $W$ minimal models'', preprint CTP TAMU-43/93, KUL-TF-93/34,
hep-th/9308114, to appear in {\sl Class. Quantum Grav.}

\item{[17]}H.G. Kausch and G.M.T. Watts, {\sl Nucl. Phys.} {\bf B354} (1991)
740

\item{[18]}R. Blumenhagen, M. Flohr, A. Kliem, W. Nahm, A. Recknagel
and R. Varnhagen, {\sl Nucl. Phys.} {\bf B361} (1991) 255.

\item{[19]}J. Figueroa-O'Farrill and S. Schrans, {\sl Phys. Lett.} {\bf
B245} (1990) 471.

\item{[20]}E. Bergshoeff, H.J. Boonstra, S. Panda and M. de Roo, ``A BRST
analysis of $W$-symmetries,'' preprint, UG-4/93'.

\item{[21]}P. Bouwknegt, J. McCarthy and K. Pilch, ``Semi-infinite
cohomology of $W$ algebras,'' USC-93/11, hep-th/9302086.

\item{[22]}E. Witten, {\sl Nucl. Phys.} {\bf B373} (1992) 187;\nl
E. Witten and B. Zwiebach, {\sl Nucl. Phys.} {\bf B377} (1992) 55.

\item{[23]}C.N. Pope, E. Sezgin, K.S. Stelle and X.J. Wang, {\sl Phys. Lett.}
{\bf B299} (1993) 247.

\item{[24]}P. Bouwknegt, J. McCarthy and K. Pilch, {\sl Comm. Math. Phys.}
{\bf 145} (1992) 541.

\item{[25]}N. Berkovits and C. Vafa, ``On the uniqueness of string theory,''
preprint, hep-th/9310129.

\item{[26]}J.M. Figueroa-O'Farrill, ``On the universal string theory,''
preprint, hep-th/9310200.

\end